\documentclass[pre,twocolumn,letterpaper,superscriptaddress,reprint]{revtex4-1}

\usepackage{graphicx}
\usepackage{pslatex}
\usepackage{amsmath}
\usepackage{booktabs, caption}

\newcommand{\atUCLA}{Dept. of Physics and Astronomy, Univ. of California, Los Angeles, Los Angeles, CA 90095.}
\newcommand{\atUH}{Dept. of Physics and Astronomy, Univ. of Hawaii, Manoa, HI 96822.}
\newcommand{\atSLAC}{SLAC National Accelerator Laboratory, Menlo Park, CA, 94025.}
\newcommand{\atCalPoly}{Physics Dept., California Polytechnic State Univ., San Luis Obispo, CA 93407.}
\begin{document}

\title{Picosecond timing of Microwave Cherenkov Impulses from High-Energy Particle Showers\\
 Using Dielectric-loaded Waveguides}

\author{P.~W.~Gorham}
\affiliation{\atUH}
\author{J.~Bynes}
\affiliation{\atUH}
\author{B.~Fox}
\affiliation{\atUH}
\author{C.~Hast}
\affiliation{\atSLAC}
\author{B.~Hill}
\affiliation{\atUH}
\author{K.~Jobe}
\affiliation{\atSLAC}
\author{C.~Miki}
\affiliation{\atUH}
\author{R.~Prechelt}
\affiliation{\atUH}
\author{B.~Rotter}
\affiliation{\atUH}
\author{D.~P.~Saltzberg}
\affiliation{\atUCLA}
\author{S.~A.~Wissel}
\affiliation{\atCalPoly}
\author{G.~S.~Varner}
\affiliation{\atUH}
\author{S.~Zekioglu}
\affiliation{\atUCLA}

\begin{abstract}
We report on the first measurements of coherent microwave impulses from high-energy particle-induced
electromagnetic showers generated via the Askaryan effect in a dielectric-loaded
waveguide. Bunches of 12.16 GeV electrons with total bunch energy of $\sim 10^3-10^4$~GeV were pre-showered
in tungsten, and then measured with WR-51 rectangular (12.6 mm by 6.3 mm)
waveguide elements loaded with solid alumina ($Al_2 O_3$) bars. In the 5-8~GHz $TE_{10}$
single-mode band determined by the presence of the dielectric in the waveguide, 
we observed band-limited microwave
impulses with amplitude proportional to bunch energy. Signals in different waveguide elements measuring
the same shower were used to estimate relative time differences with 2.3~picosecond precision. 
These measurements establish a basis for using arrays of alumina-loaded waveguide elements, with exceptional
radiation hardness, as very high precision timing planes for high-energy physics detectors.  
\end{abstract}
\maketitle

\section{Introduction}

Future colliders with center-of-mass energies in the tens to even a hundred TeV are now
under detailed study~\cite{TeVColliders,Chekanov17}. Among the most pressing design issues is the need for
new methods to address the daunting levels of pileup in
detectors near the collision region. Equally challenging are the high levels of ionizing radiation
near the beam, leading to more and more rapid degradation of detector elements as luminosity grows.
One of the most promising ways to make progress on the former issue
is to improve the detector timing precision down from the nanosecond to the picosecond level. And
in the case of the latter issue, extreme radiation environments will require use of materials with intrinsic
radiation hardness. 

Significant effort is now underway to improve the timing of current collider detector technologies,
and timing resolutions of 15-30 ps with silicon-based detectors 
have been demonstrated in some cases~\cite{White14,Albrow14,Cartiglia17,Apresyan17}.
Traditional vertexing of collider events has been primarily three-dimensional, with the event
time generally known to no better than 100~ps, equivalent to 30~mm of spatial precision, far worse than
the sub-millimeter precision of spatial trackers. In fact because of the finite duration of
bunch crossings in the collision region, events may share the same vertex but a different time,
or occur at the same time, at different vertices. With pileup events projected to be in the hundreds
for the high-luminosity upgrade to the LHC, the need for four-dimensional vertexing, including precision
timing, is becoming acute. Planned timing upgrades to the 20-30~ps range will lead to 
timing constraints of order 1~cm, a significant improvement but still an order of magnitude from
the spatial tracker constraints. It is evident that for future colliders at even higher collision energies,
fully commensurate four-dimensional fits to vertex positions
will become critical to optimizing performance of these systems.

With such precision timing and radiation hardness goals in mind, we have performed an experiment to test the possibility
that coherent microwave Cherenkov signals arising from the Askaryan effect~\cite{Ask62} can be used to
for precision characterization of high-energy particles via their secondary electromagnetic showers.
To achieve adequate sensitivity, we use cryogenic cooling to reduce thermal noise, 
and dielectric-loaded waveguide elements to yield controlled geometries that afford high-precision timing.
Copper waveguide loaded with solid alumina, among the most radiation-hard dielectrics known, 
is used as the detector element. Our goal was to establish scaling relations for the least-count energy
in such a detector, and how arrays of such elements might perform as a timing instruments for
constraining vertex geometry in a sampling high-energy physics detector.
These goals have been investigated via the SLAC National Accelerator Laboratory
program, experiment T-530, entitled the Askaryan Calorimeter Experiment (ACE)\cite{Note1}.

The Askaryan effect -- coherent radio Cherenkov emission from the negative charge excess in
a high-energy electromagnetic cascade~\cite{Ask62} -- was first confirmed at SLAC in 2001~\cite{SLAC01}
using a silica-sand dielectric target in the T-430 experiment. Cosmic-particle-induced radio impulses based
on the Askaryan effect provide the basis for a host of particle astrophysics experiments within the last
two decades primarily aimed at detecting ultra-high-energy neutrinos~\cite{Hankins, GLUE, RICE, ANITA}. 
In these experiments, the high-energy particle cascades take place in unbounded
dielectrics, such as cold polar ice, or the lunar regolith, and the resulting radio signals are detected by
embedded or external detectors. In general, because of the large scale of these experiments and the presence
of natural thermal noise, the primary particle energy threshold for the radio emission to exceed 
thermal noise is in the tens of PeV to EeV range or more. Such methods thus have to date not been viewed
as relevant for precision particle detection in high-energy physics accelerator or collider experiments.

Coherent microwave to millimeter wave emission from high-energy charged particle bunches has
been studied at accelerators for many years, primarily with a view to improve beam diagnostic methods.
Takahashi {\it et al.} (2000)~\cite{Tak2000} specifically studied the generation of mm-wave coherent Cherenkov
radiation from electron bunches in close proximity to, though without entering, adjacent dielectrics.
Other sources of coherent millimeter and microwave radiation including synchrotron, transition radiation, 
and Smith-Purcell radiation have also been studied under similar conditions~\cite{cohRad1,cohRad2}. In these experiments, the
bunch charges are typically very large, to enhance the microwave or mm-wave power by the quadratic dependence
of power on bunch charge in coherent radiation. To date, these investigations have not considered
detection of particles using secondary emission from the Askaryan effect, but the theory developed in support
of these experiments will help to provide a context for evaluating Askaryan-induced emission 
as a high-energy physics tool.

In the following, section II outlines the theory of Cherenkov emission from a finite track in a dielectric.
Section III provides a detailed description of the detector design, including (A) GEANT4 simulations 
of the particle showers in the detector, (B) finite-different-time-domain simulations of the 
electrodynamics of this process, and (C) applications of these results to the theory of section II.
Section IV describes our beam test experiment and the results, which include 
(A) high and (B) low beam current tests, (C) energy calorimetric results, and (D) fast timing results.
Section V discusses some important issues such as (A) the effects of system noise temperature,
(B) track length in the detector, (C) magnetic field effects, and concludes with (D) some
discussion of applications for the methodology.

\section{Background theory}

Our detector concept employs a finite track of a relativistic charged particle bunch through a dielectric
bounded on either end of the track by the conductive waveguide wall, 
a classical application of Cherenkov radiation
from finite tracks,  first described by Tamm in
1939~\cite{Tamm39}. The charged bunch can originate as a secondary shower from a single
primary energetic particle, in which case the charge excess is the relevant contributor, 
or it could be a pre-collimated bunch delivered by an accelerator. 
The coherent microwave and millimeter emission from the latter process
was studied in detail in the 1990's in accelerator experiments~\cite{Tak2000}. In these efforts
the difference between Cherenkov and transition radiation becomes indistinct when the
track length becomes comparable to the wavelength scale of the frequencies involved,
as is the case for our experiment. 

Tamm's theory did not distinguish between the nature of the radiation by this nomenclature, and
it applies to either case. The theory is derived for only emission from a single charged
particle, and thus coherent emission from many charges partially or fully in-phase must include
the appropriate phase factors in computing their radiation field. Tamm also assumed 
open boundary conditions for the radiation; thus a bounded waveguide structure will also modify the results.
Although these modifications lessen the direct applicability of Tamm's theory to our
experiment, we will use it as a basis for comparison, particularly because it encompasses
both Cherenkov and closely-related transition radiation which are the appropriate emission
processes in our case.

For a finite-length charged particle track in a dielectric bounded by conductors, the power per unit
solid angle $d\Omega$ , per unit frequency interval $df$ is
\begin{equation}
\frac{d^2 W}{d\Omega df} ~=~  -\frac{\alpha n f L}{c^2} \left ( \frac{\sin X(f,\theta)}{X(f,\theta)} \right )^2 \sin^2 \theta
\label{eq1}
\end{equation}
with the frequency-dependent angular response $X(f,\theta)$ given by
\begin{equation}
X(f,\theta)  = \frac{\pi L f}{\beta c} ( 1 - \beta n \cos\theta)~.
\end{equation}
Here $\alpha$ is the fine structure constant, $n$ the index of refraction of the dielectric, $L$ the path length,
$\beta = v/c$ the normalized particle speed, and $\theta$ the polar angle relative to the track.
For a number $N$ of co-propagating charged particles in this scenario, the resulting power will depend 
on the summed electric fields produced by each particle, with respective phase factors:
\begin{equation}
\mathbf{E}_{tot} ~=~ \sum_{j=1}^N \mathbf{E}_j \exp \left ( \frac{2\pi i f}{c} \hat{r}\cdot \mathbf{x}_j \right )~.
\label{formfactor-eq}
\end{equation}
Here $\mathbf{E}_j$ is the field from the $j$th particle,  
$\mathbf{x}_j$ is its position, and $\hat{r}$ the unit vector in the direction of observation.
We will use this formalism for a first-order analysis of the expected signal in a later section.

\section{Detector Design}

A high-energy charged particle or jet of particles propagating out from a collision vertex will generate
showers along its track in any non-vacuum portion of a detector system. Of course measurements
of these showers are essential to calorimetric detectors. The material in which the shower is generated
must therefore be able to either directly sample the shower particle density, or measure it by
secondary emission. In our case, while sampling of showers using radio Cherenkov emission is calorimetric,
its relatively high least-count energy resolution (at least in the current realization) leads to
reduced importance of the calorimetry in favor of the timing resolution. 

Whether used for calorimetry or timing, the
material that samples the shower must be transparent to the microwave emission generated in it, as 
this emission must propagate over a significant distance in the waveguide element. Since
this material is likely to also be very near the interaction region, especially in a forward detector,
it must also possess a high immunity to radiation damage. Materials of higher density
and higher microwave dielectric constant will also increase the coupling of the microwave signal
to the waveguide element. Finally, the manufacturability of the material into shapes commensurate
with a rectangular waveguide is important. 
These considerations lead to alumina, a ceramic available at low cost with very high purity,
as the material of choice in our detector. 

For alumina (aluminum oxide, Al$_2$O$_3$) at microwave frequencies, the real part of the relative
dielectric constant is typically $\epsilon_r \simeq 10$ and remains constant over a wide range of frequencies.
The dielectric loss tangent of pure Alumina at microwave frequencies is typically $\tan{\delta}< 2-3 \times 10^{-5}$, 
among the lowest of any material known; only sapphire (which is also a form of Aluminum oxide) and
fused silica possess lower loss tangents, at least among any commonly available solid dielectrics.
Fused silica is also an excellent choice from the point of view of its low loss tangent and radiation
hardness, however its density and dielectric constant are of order half that of Alumina and sapphire,
which has impact both on the frequency response of loaded waveguide elements, and in reducing the
coherent Cherenkov signal.
Synthetic sapphire exceeds Alumina in all relevant microwave and material properties, but would require
a costly manufacturing process for rectangular bars. 

\begin{figure}[htb!]
 \includegraphics[width=3.5in]{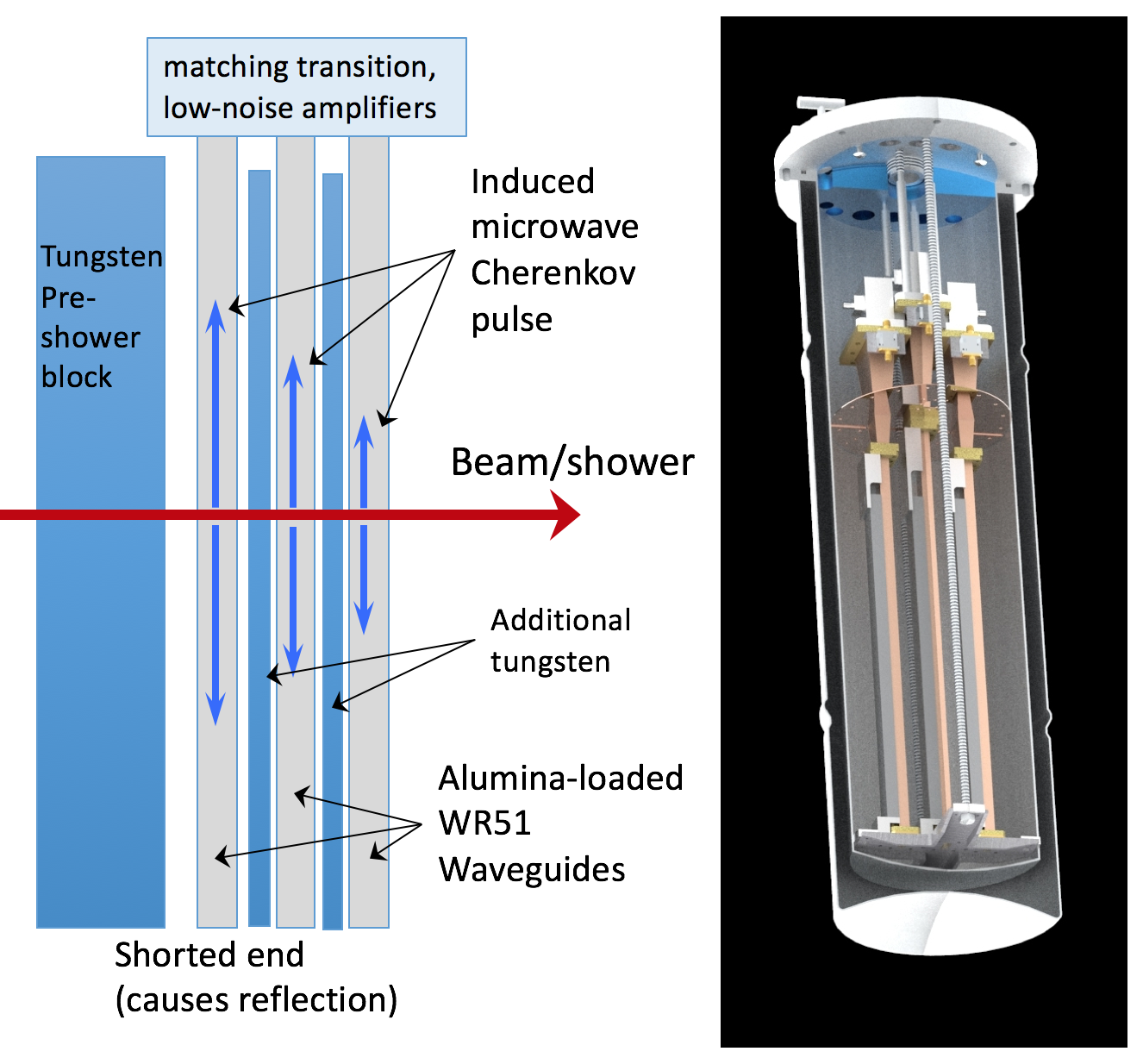}
 \caption{ Left: Block diagram of basic geometry of the detector. Right: Rendering of the detector elements
 inside of the liquid nitrogen dewar.
 \label{diagram}}
 \end{figure}

Alumina is used ubiquitously for electrical insulators in a very wide range of applications, including in
reactor cores where the radiation environment (particularly neutrons) is very intense. It is also among the
best and most robust of high-voltage insulators; 
automobile spark plugs are commonly built with Alumina insulators.
Under heavy irradiation, Alumina can develop slight radiation-induced 
conductivity (RIC)~\cite{Al1,Warman91}. This RIC never reaches a level capable of causing
high voltage breakdown, and the impact on the microwave loss tangent is also small~\cite{Molla95}.

Alumina is also quite dense among standard ceramics, $\rho = 3.5$~g~cm$^{-2}$. This leads to more
compact shower development within the material, and a smaller Moliere radius, leading to microwave
coherence at higher-frequencies than other low-loss dielectrics such as polyethylene, teflon, or fused silica.

These considerations, as well as the ready availability of high-purity rectangular alumina bars that fill standard
waveguide shapes, led us to choose alumina as the dielectric load in our beam test waveguide elements.
For rectangular waveguide uniformly loaded with a low-loss, non-magnetic dielectric, the 
frequency of the waveguide cutoff for the lowest-order $TE_{10}$ mode scales as $\epsilon_r^{-1/2}$:
\begin{equation}
\omega_c = \frac{c \pi}{\sqrt{\epsilon_r} a}
\end{equation}
where $a$ is the larger dimension of the rectangular waveguide, $c$ is the speed of light.
The upper limit for single-mode operation depends on the appearance of the next-order mode;
for standard rectangular waveguide, this limits the upper range to $\omega_u \leq 2\omega_c$.
Operation too close to the lower frequency cutoff is also difficult because of dispersion effects,
so one typically moves up about 15-20\% above the cutoff to ensure a clean signal transmission.

Thus for WR-51, with cross-sectional inner dimensions of $12.96 \times 6.48$~mm,  the normal
waveguide cutoff for the unloaded $TE_{10}$ mode is 12~GHz, but due to impedance dispersion close to the
cutoff, the practical operating single-mode bandwidth  is 15-24~GHz.
The operating band moves down to 5-8~GHz when loaded with Alumina. For standard waveguide this poses
another problem: the intrinsic impedance also scales in a similar way, and thus the nominal 50~$\Omega$
waveguide develops a characteristic impedance of closer to 17~$\Omega$. An impedance
transition is therefore necessary to efficiently couple out the signals into standard $50~\Omega$ coaxial
cable.

Fig.~\ref{diagram} shows both a conceptual diagram (left) and solid model (right) 
of our initial experimental prototype. Three 
loaded waveguides are aligned transverse to the accelerator beam. A $3.6$ radiation length (RL) 
tungsten-alloy bar (90\% W, 7\% Ni, 3\% Fe) is used as necessary to 
pre-shower the particle bunch, and additional 2.5~RL tungsten bars also may be used 
to further increase the rate of shower development between 
the waveguide elements. The combined radiation length of the two 1~mm copper waveguide walls, and
the 6.3 mm thick Alumina bar contributes and additional 0.23 radiation lengths per ACE element.
Microwave emission is induced within the waveguides by the $e^-  - e^+$ charge excess, coupling
to the internal waveguide modes, and then propagating to both ends. On one end, the received signal passes through
an impedance-matching transition, and is coupled out to a microwave low-noise amplifier (LNA). On
the other end, the waveguide is shorted, and the signal reflects as an inverted pulse, and then
propagates back up to the LNA. 

The charge excess can also include a significant or even
dominant contribution
from the input charge of the bunch.  In fact we used bunches with high charge levels,
and no tungsten elements, to initially verify the system performance. 
Once the input bunch charge falls below several hundred electrons,
the secondary charge of the EM shower becomes dominant, and the Askaryan charge excess
supplies the net current element that induces the microwave impulse.

We then use secondary amplifiers to increase the signal up to a level
where it is then able to be digitized using a Tektronix TDS6804 (8 GHz bandwidth) 
or TDS6154 (15 GHz bandwidth) oscilloscope. The right side of Fig.~\ref{diagram} shows a rendering of the actual
detector assembly, contained within a liquid nitrogen dewar. 

 \begin{figure*}[htb!]
 \includegraphics[width=7in]{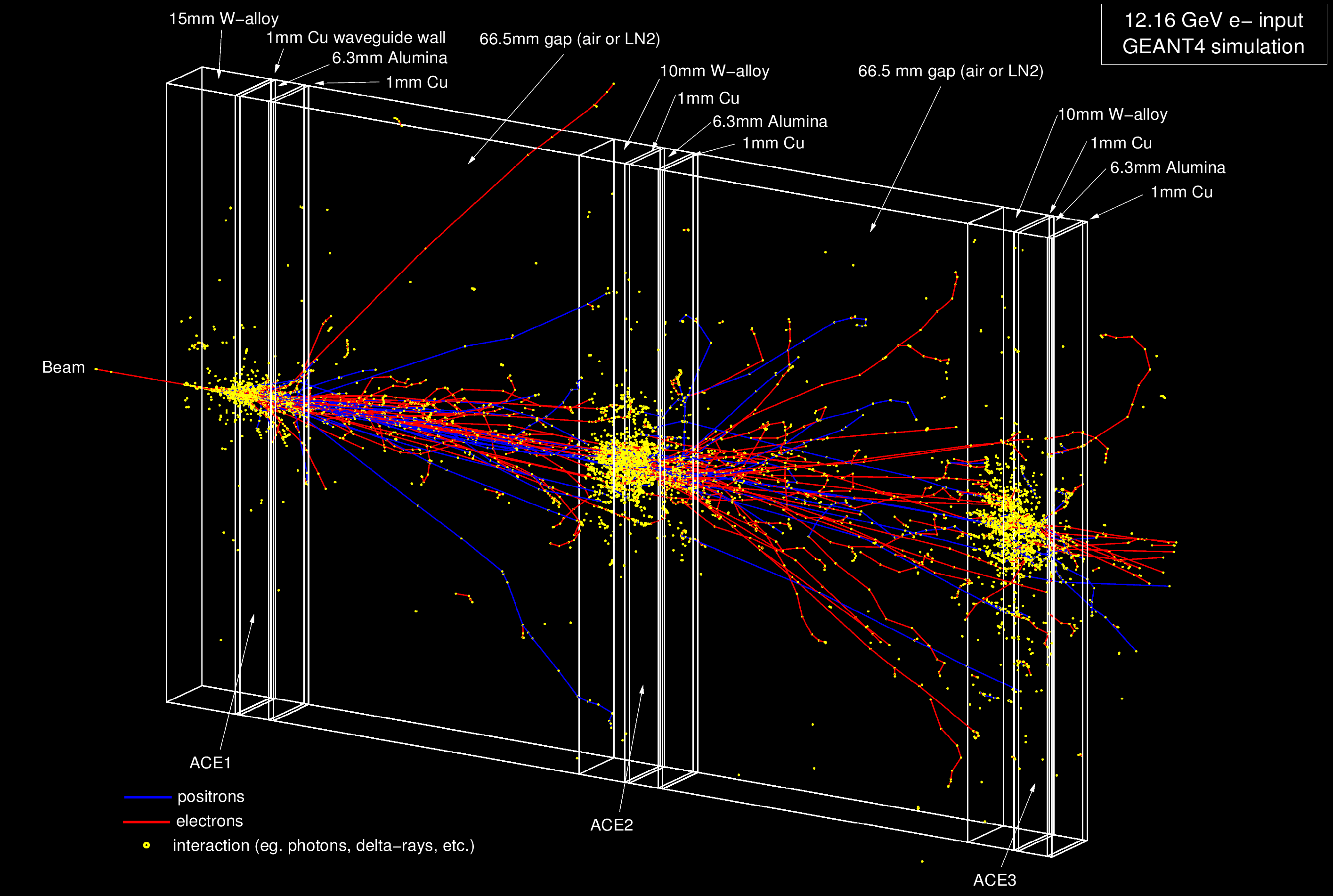}
 \caption{Layout diagram, and GEANT4 simulation of a single 12.16 GeV electron event in our ACE detector system;
 in this case liquid nitrogen occupies the interelement spaces.
 \label{geant4a}}
 \end{figure*}

\subsection{GEANT4 shower simulations}

  \begin{figure*}[htb!]
 \includegraphics[width=6.5in]{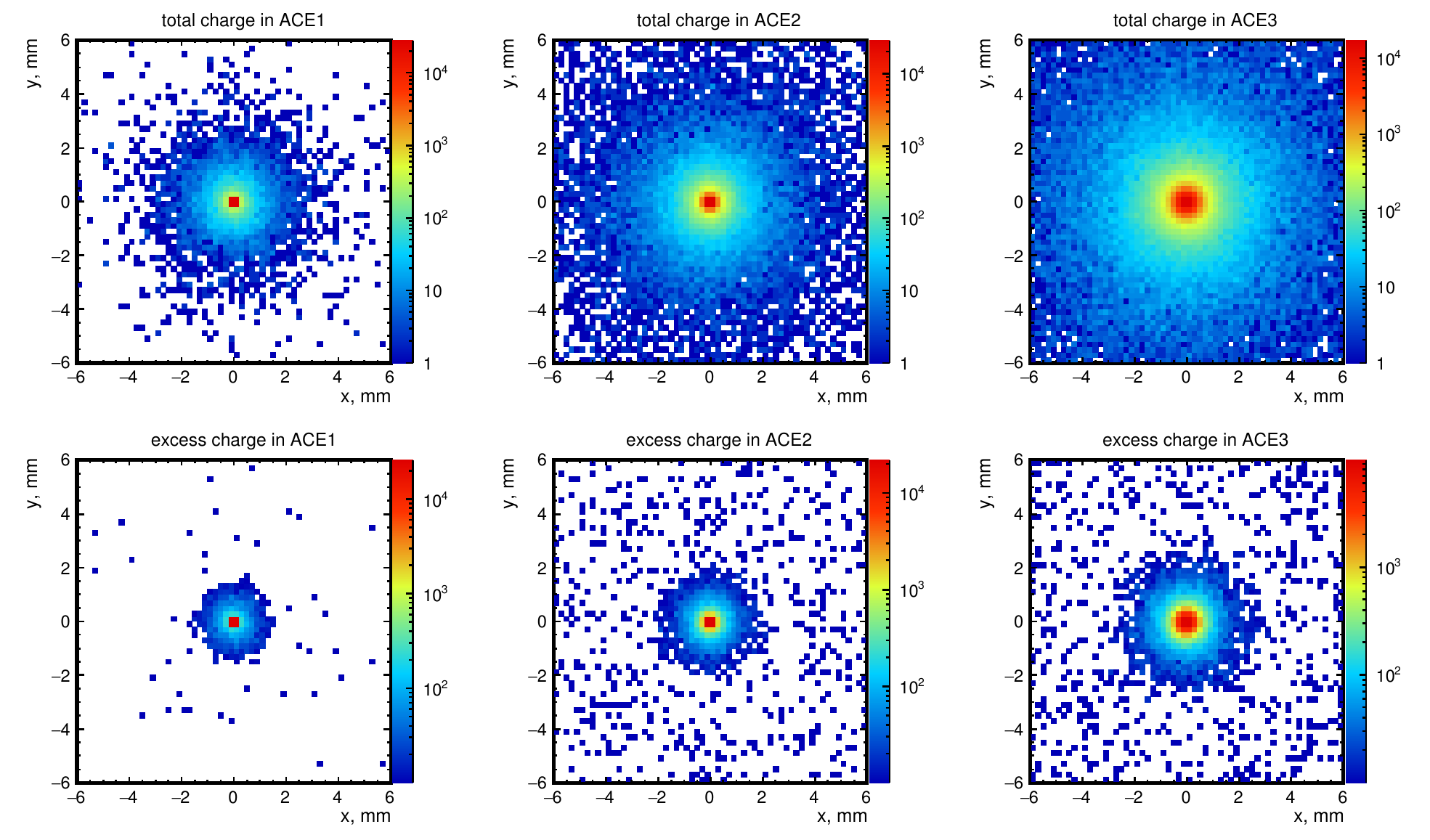}\\
 \includegraphics[width=6.5in]{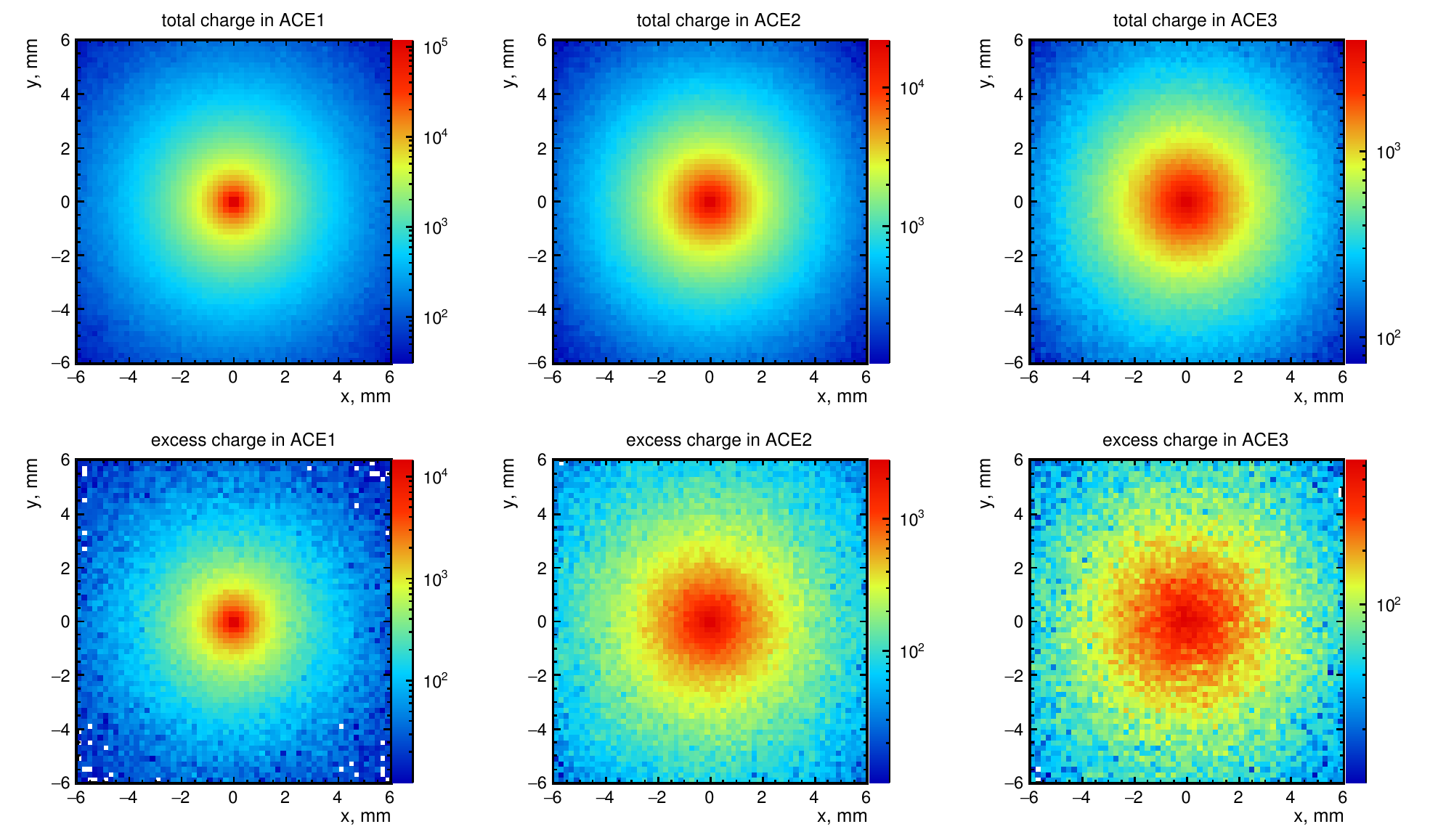}
 \caption{GEANT4 simulation of total and excess charge in the 3 ACE elements, averaged over $10^5$ events.
 The top 2 rows give the results for 6.4 GeV electrons, with no tungsten in the ACE system; only slight evolution
 of the bunch can be seen. The bottom two
 rows are for the full tungsten-loaded ACE with 12.16 GeV electrons; strong evolution of the shower is
 now evident.
 \label{geant4b}}
 \end{figure*}

Even relatively simple multilayer particle detectors produce showers with a high degree 
of complexity due to the myriad of possible interactions. To simulate the
shower development in ACE we created a GEANT4 model
of our system which allows us to predict the total and net charge distributions in
each ACE element, along with time and spatial profiles to evaluate the form factors of
the evolving shower front as it passes through the system~\cite{GEANT}. 

Fig.~\ref{geant4a} shows a labeled wireframe representation of our detector, along with
an example of a single simulated shower from a 12.16 GeV electron. For simplicity we
have ignored the beam pipe exit window (very thin aluminum), the $\sim 3$~m of air between
this exit and our dewar, and the $\sim 2$~mm of stainless steel in the dewar wall. Positrons are
shown as blue tracks, electrons as red tracks. Yellow dots mark vertices of a variety of interactions and collisions,
and here we include those vertices from photons, although the photon tracks themselves
are suppressed for clarity. Because Alumina has a high microwave index of refraction $n=3.15$,
and we are interested in all $e^+e^-$ that could produce Cherenkov emission, we
do not cut off any low energy electrons, down to well below the Cherenkov threshold determined
by $\beta n=1$.

 \begin{figure}[htb!]
 \includegraphics[width=3.5in]{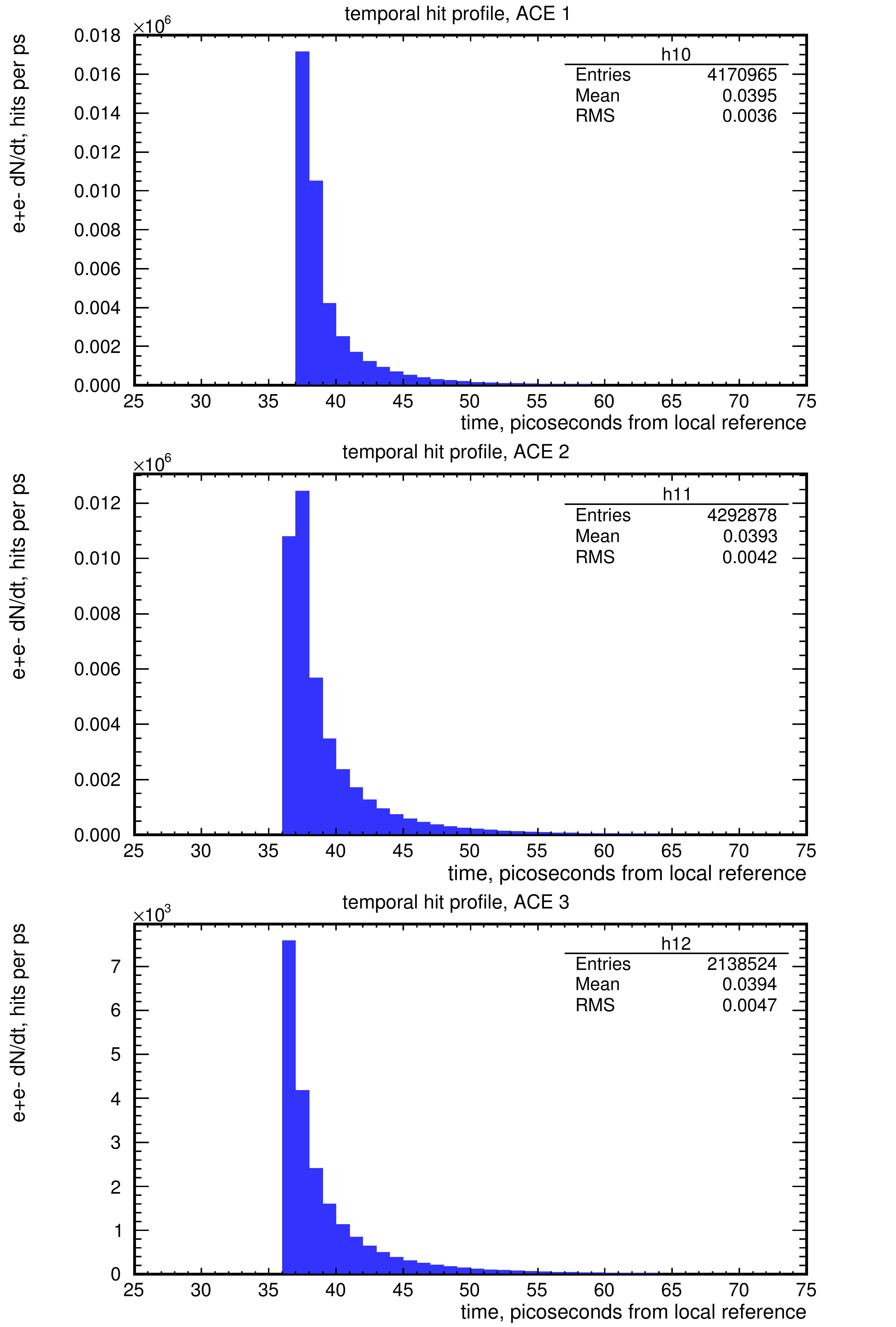}
 \caption{GEANT4 simulation time profiles of total charge in the 3 ACE elements, averaged over $10^5$ events.
 This is for the case of 12.16 GeV showers and a full set of tungsten absorbers present. 
 \label{geant4c}}
 \end{figure}
 
Fig.~\ref{geant4b} shows the simulated distribution of charge passing through a $12.6 \times 12.6$~mm
portion of the detector around the beam axis for two cases: the upper two rows for 6.4 GeV electrons and
no tungsten, and the lower two rows for 12.16 GeV electrons and the full load of tungsten, including the
pre-shower block and additional blocks between the elements. In each
case $10^5$ electrons were simulated.
Shower development in the first case is observed to be very limited, consistent with the 0.23 radiation length
of material per ACE element. To compute the charge excess, we assign opposite weights to positrons (-1) and
electrons (+1) in the histogram, which is clipped at zero. In the second case, we expect shower maximum to
occur in the tungsten prior to the second ACE detector, and thus the shower is declining by the third detector.

For purposes of estimating the net charge for Tamm's theory, we integrate the charge (or charge excess)
within the inner quarter-wavelength central region around the beam axis, since beyond that limit
the Rayleigh criterion will no longer be satisfied and destructive interference will diminish
the coherent field contribution. A better approach would be to couple the GEANT4 simulations
directly to an electromagnetic simulation, using equation~\ref{formfactor-eq} above to determine the
phase factors, but this option is complex to develop for our geometry
and thus beyond our scope in this report. 

The average charge excess in each of the three detector elements from Fig.~\ref{geant4b}
is integrated over a centered $6.3 \times 6.3$~mm area and scaled to the
equivalent of 100 incident beam electrons, giving a net excess charge of 112, 117, and 122 electrons per 100 incident
electrons for ACE1, ACE2, and ACE3 in the case with 6.4 GeV bunches and no tungsten. For the
12.16 GeV case with full tungsten, the excess charge is 611, 493, and 199 electrons per 100 incident electrons,
corresponding to 18.2, 19.0, and 20.6\% of the total charge
for ACE1, ACE2, and ACE3 respectively. Since the amplitude of signals generated via the Askaryan effect
is proportional to the coherent charge excess, we can expect that the amplitudes observed in our 
detectors should follow these ratios to first order. In addition, we can use these net charges as inputs to Tamm's 
theoretical estimates of the microwave emission power.

In addition to the effect of space charge localization on microwave pulse coherence, we also consider the
time coherence, since particles scattered away from the primary beam axis may also suffer delays.
Fig.~\ref{geant4c} shows the arrival times of the shower particles in each ACE element for the fully
loaded tungsten case with 12.16 GeV electrons. In the temporal domain, the highest single-mode frequency
in our system is 8~Ghz, giving 125 ps per cycle, or just over 30 ps per quarter cycle. It is evident that
the vast majority of the charge in each case arrives well within 10 ps of the lead bunch arrival time, and
we thus expect the time delays to have negligible effects on the phase coherence of the signals.

\subsection{Electromagnetic signal simulation for initial design}

 \begin{figure*}[htb!]
 \includegraphics[width=7.2in]{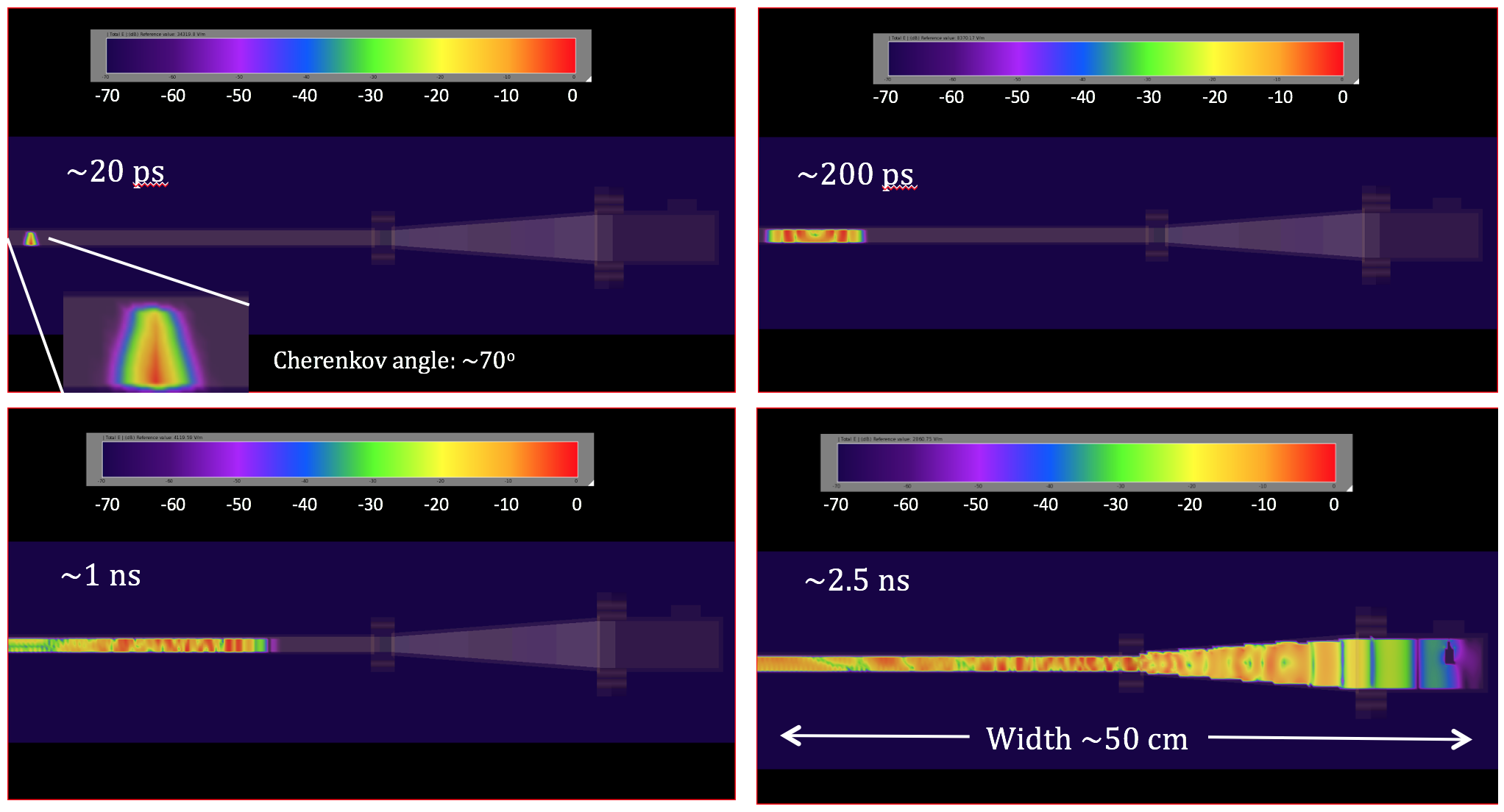}
 \caption{Four discrete time snapshots of a cross section through one of the simulated ACE
 loaded waveguide elements, indexed by the time relative to the beam entry through the 
 element. The beam transits from bottom to top. Amplitude is indicated
by the color scale in dB relative to the peak amplitude. 
 \label{acesim1}}
 \end{figure*}

There is considerable uncertainty in adapting current open-boundary condition theory 
and semi-empirical analysis for coherent microwave emission from an EM shower to the conditions 
of a closed dielectric-loaded waveguide. As noted above, the sudden appearance and disappearance of the
charge excess as the shower enters and exits the waveguide will produce transition radiation that
may constructively or destructively interfere with the coherent Cherenkov emission. In addition,
the coupling of the radiation to the waveguide modes takes place in near-field conditions where
calculations are challenging. Both Cherenkov and transition radiation produce emission that is
azimuthally symmetric around the charged bunch track, but only a portion of this emission can
couple to the longitudinally-propagating $TE$ and $TM$ modes of the waveguide. And finally,
the very sharp impulsive nature of the emission produced by these radiation mechanisms does
not lend itself well to standard single-frequency, steady-state analysis methods. While Tamm's
theory does account for the combination of TR and CR, it does not provide guidance on how
to estimate the instrinsic time-domain pulse shape of the signal, which is critical to our
detection process.

Fortunately there are numerical simulation methods developed for time-domain analysis which inherently
obey Maxwell's equations and are capable of high precision even in very complex geometries and
with unusual source stimuli. These Finite Difference Time Domain (FDTD) methods
provide a framework for gridded EM simulations that can be used for our application, with some
careful attention to the representation of the source current, which in our case is a relativistic current impulse.
For our simulations, we build models using Remcom's XFDTD simulator~\cite{XF7}.


Fig.~\ref{acesim1} shows four time slices of the signal generated in one simulation through
one of the detector elements. The beam transits from bottom to top in the initial, upper left pane, and generates a
transient signal with Cherenkov-like characteristics, at an angle of $\sim 70^{\circ}$ relative
to the beam direction, consistent with the radio index of refraction of alumina. In the second frame of the sequence, the
signal can be observed to propagate symmetrically in opposite directions within the waveguide, although in these
frames we only follow one of the signals to the waveguide transition.
In the third (lower left) frame, the leading portion of the pulse couples to the dominant $TE_{10}$ mode in the waveguide, while
some trailing transient modes are still evident. In the final frame, the wavelength
then expands as the signal passes into the tapered transition and then into the waveguide-to-coaxial
adapter, through a stepped gradient of the index of refraction.

 \begin{figure}[htb!]
 \includegraphics[width=3.5in]{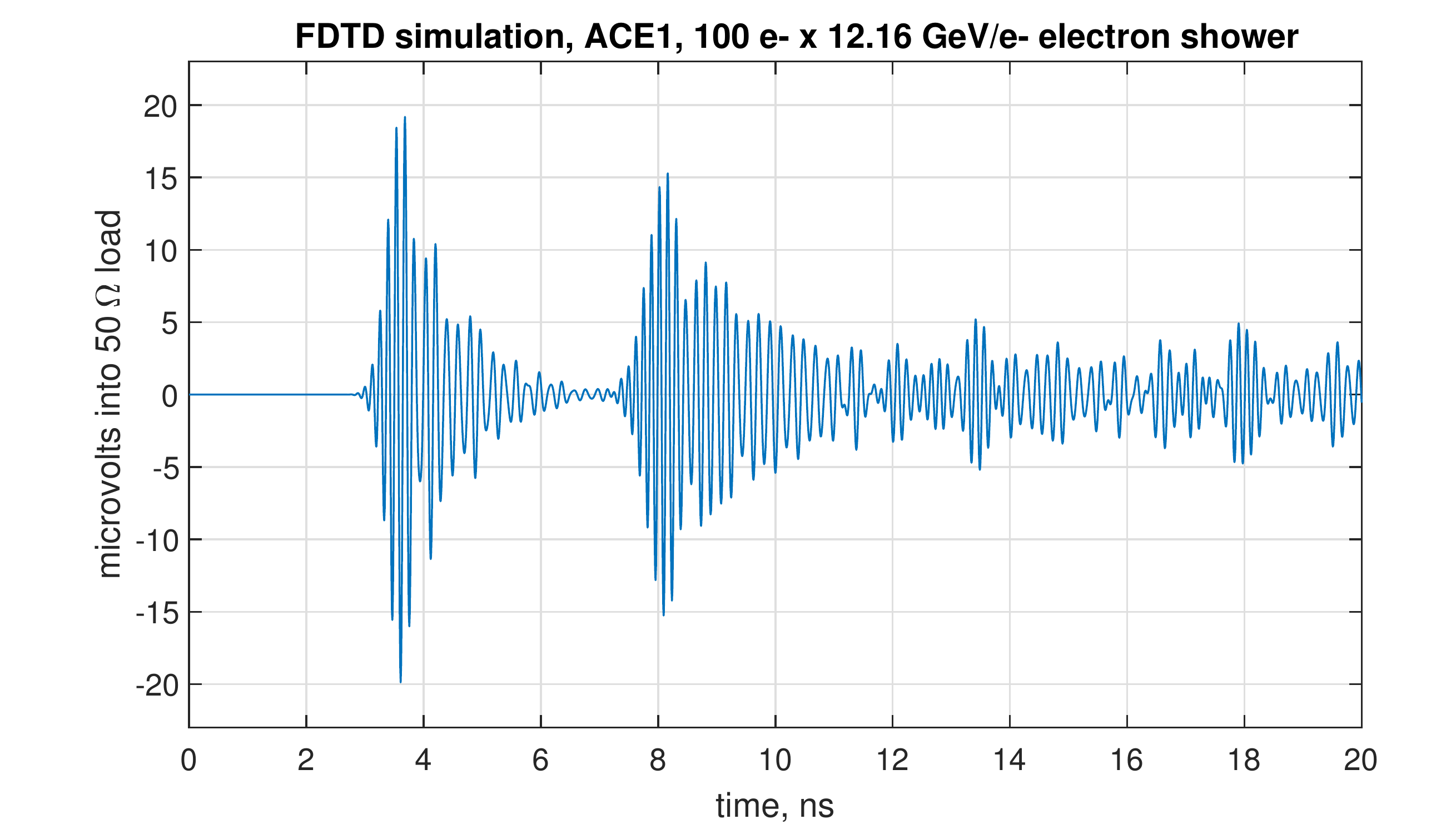}
 \caption{ 
 FDTD simulation of as-built ACE element response to a transverse shower
 of 100 12.16 GeV electrons, entering after $\sim 4$ radiation lengths of
 tungsten. The distal end of the waveguide has been shorted, producing a
 reflection as seen in the waveform.
 \label{ACEXF}}
 \end{figure}


Fig.~\ref{ACEXF} shows a detailed FDTD simulation of the as-built ACE detector, excited via a source designed
to closely mimic the current impulse produced by a relativistic charge bunch transiting the waveguide element.
In this case one end of the waveguide element has been shorted as in the deployed version of the detector.
The current pulse amplitude matches the GEANT4-determined amplitude for a 100-electron 12.16 GeV shower
through the initial tungsten radiator, giving a 611-electron bunch in the waveguide element. The resulting
response includes the reflection from the far end, as well as the residual ringing caused by the lack of
matched termination of that shorted end. The pulse resulting from the secondary reflection also transits a much longer section of
the waveguide, leading to a larger degree of absorption and waveguide dispersion as observed. 

This waveform has also been
filtered to limit the output to only the 5-8~GHz single-mode response of the waveguide; this is necessary
since the very sharp current impulse excites all of the higher-order modes of the waveguide as well, but
these are generally too complex in their behavior to be of use for these measurements.
We do not show the expected thermal noise of the system in this simulation; but we note that
the $\sim 20\mu$V peak output of the leading pulse is about 3.3 times the thermal noise level
at the $\sim 18$K system temperature measured for the detector when cooled by liquid nitrogen.

The FDTD simulations illustrate
one of the features that enables very precise timing: the modulation pattern observed
provides a very effective time-domain vernier with frequency content much higher than is
typical for detectors used in high-energy physics. This is due in part to the effects of waveguide
dispersion, which extend the duration of the pulse and increase the number of cycles of the
passband modulation. This comes at some expense in the overall peak amplitude of the pulse,
but because it is deterministic depending on the waveguide used, it can be removed completely 
in analysis.

\subsection{Tamm theory estimation}

The FDTD results, which rely only on a finite-element implementation of Maxwell's equations, include all of the 
electrodynamic radiation mechanisms for a relativistic charge bunch passing through a loaded waveguide,
and also include the intrinsic waveguide coupling, dispersion, along with inefficiency and losses.
It is howeve useful to also make a signal strength estimate based on
Tamm theory to confirm the order-of-magnitude of the simulation.

\begin{figure}[htb!]
 \includegraphics[width=3.35in]{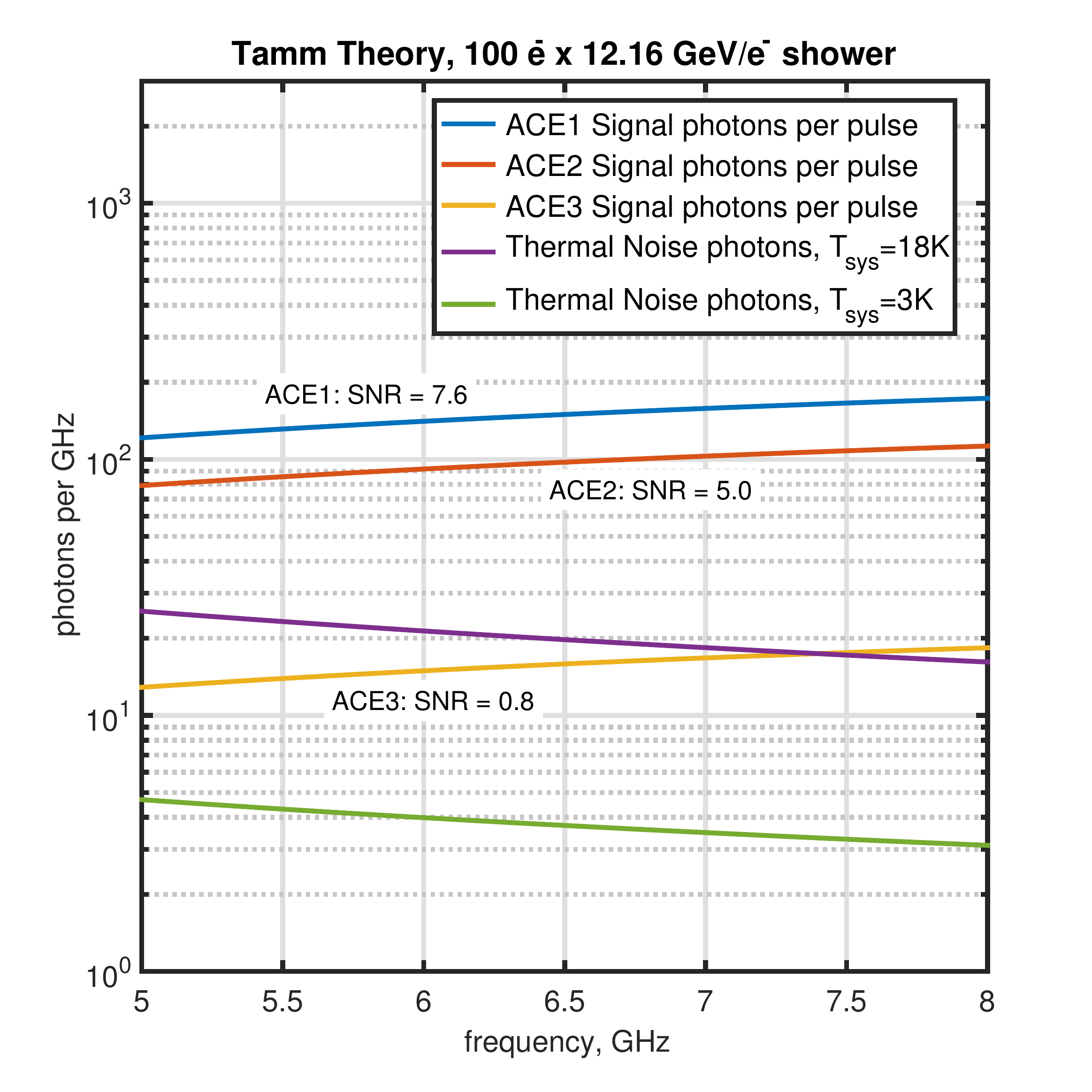}
 \caption{ Prediction for our experimental parameters from Tamm theory.
 \label{Tamm1}}
 \end{figure}
 
To do this, we introduce an {\it ansatz} regarding the coupling efficiency of the finite-length Cherenkov emission to
the waveguide. We assume that energy emitted within $\pm 30^{\circ}$ around the longitudinal axis of the 
waveguide will be coupled into the lowest order mode, and we numerically integrate 
equation~\ref{eq1} above over this range. This assumption is motivated by standard methods
of coaxial-to-waveguide coupling, which make use of quarter-wave monopole feeds to excite
the $TE_{10}$ mode in standard rectangular waveguide. 

Fig.~\ref{Tamm1} shows the results of this
for the parameters of our system, using alumina-loaded WR-51 with a system noise temperature
we estimated at $T_{sys} =18$~K, and a bunch charge equivalent to that produced by a 1200~GeV shower via
the Askaryan effect. The system temperature is dominated by the low-noise amplifier; since the 
alumina loss tangent is negligible it does not contribute to the thermal noise, and the waveguide
losses at our frequency range also contribute only a fraction of a Kelvin of thermal noise.
The results of Fig.~\ref{Tamm1} are given in terms of differential photon counts for both
the signal and the background thermal noise. We also include the thermal noise for a
much lower noise commerical amplifier, giving $T_{sys}=3$~K. Because 
we are operating with very low-noise amplifiers and in cryogens, the resulting signals
approach the regime $h\nu \sim kT_{sys}$, and we thus use the 
proper estimator for the standard deviation $\sigma_N$
in photon counts~\cite{Zm03}
\begin{equation}
 \sigma_N = \frac{1}{\eta} \sqrt{\eta N(1+\eta N)}
\end{equation}
where $\eta$ is the receiver efficiency (about 0.9 in our case), and $N$ is the mean
photon count integrated over the frequency band. Fig.~\ref{Tamm1} also indicates the
signal-to-noise ratio for each of the three ACE channels in the fully loaded
tungsten case. 

The results indicate a SNR which is about a factor of 2 higher than that given by
the FDTD simulation. It is likely that several factors, including lower coupling 
efficiency than we have assumed, along with waveguide losses and dispersion, contribute
to a reduction of the peak amplitude in the FDTD simulation, and accounting for these
factors the two methods give reasonable agreement. In fact, as we will show in later analysis,
deconvolution of the waveguide response using correlation methods recovers most
of the difference in SNR that we observe here.


\section{Beam Test Description \& Results}

Our initial ACE beam test, was performed
at the SLAC National Accelerator Laboratory as experiment T-530, and took place over
the period August 5-10, 2015 in the End Station A Test Beam facility (ESTB).

Because the beam parameters were set by experiments underway at the 
SLAC Linear Collider Light Source (LCLS) facility, we
initially ran with electrons of energy ranging from 4-6~GeV. Beam currents were controllable
from $10^9$ electrons per bunch down to as low as several tens of electrons per bunch.
The gain of our low-noise amplifiers and signal chain, when fully instrumented,
was of order 65 dB, allowing us to reach thermal noise levels for the low-beam current
cases. Initially, however, to establish timing and optimize the placement of our detectors,
we ran at high beam currents without amplification. 

Since the generation of microwave Cherenkov emission
arises from any beam current passing through our detector, whether it is part of the charge excess in an EM shower,
or just the current impulse from a transiting electron bunch, our initial testing was
done without the tungsten in place. In fact this was preferable during the runs with
4-6 GeV bunch energy, since at this energy, the multiplication of charge in the shower
is much reduced compared to higher energies, thus reducing the contribution from the Askaryan
effect compared to the direct Cherenkov of the transiting bunch. These
earlier runs allowed us to optimize our configuration for the final testing which was
done with 12.16~GeV bunches, for which the Askaryan contribution is dominant.

\subsection{High beam current tests}

\begin{figure}[htb!]
 \includegraphics[width=3.5in]{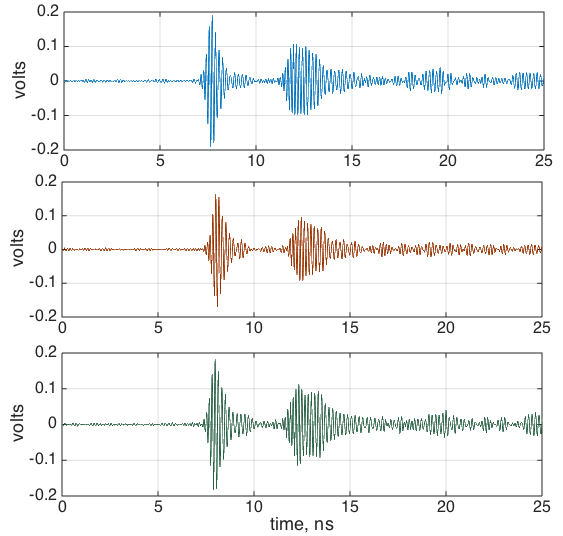}
 \caption{ ACE data using high-beam current runs, with $\sim 10^8$ electrons per bunch.
 No amplifiers or pre-shower tungsten blocks were used in this configuration.
 Top: ACE1, Middle: ACE2, Bottom: ACE3. Both the primary pulse and the reflection
 from the shorted waveguide end are evident; the reflected signal is broadened
 due to waveguide dispersion.
 \label{ACEday1}}
 \end{figure}

Fig.~\ref{ACEday1} shows a typical set of beam-induced impulses from a run with a 6~GeV beam,
taken with a high beam current of $\sim 1.2 \times 10^8$ electrons per bunch, 
and no LNA amplification or tungsten radiators. The goal for these initial data was to establish
the behavior of the underlying emission process in the absence of thermal noise or significant shower
development. The figure panes from top to bottom are the microwave signals
for the three WR-51 elements from front to back. No correction for cable timing offsets have been
done in these data, thus the time delays in the traces do not yet accurately reflect the physical
offsets. As noted above for the GEANT simulations, each ACE element is only 0.23 radiation lengths thick,
and thus shower development between
the first and third element is only slight, leading to similar signals in all three detectors. 
Variations are due to slight differences in the coupling and also the length of the three
elements; the center element was 2.5~cm for mechanical reasons.

The initial peak in each of the traces in Fig.~\ref{ACEday1} is the direct signal from the
beam transit, followed by the reflection off the far end, which suffers from roughly three times
the dispersion of the direct signal, and is thus reduced in amplitude. Comparison of the
shape and modulation of the direct peak to the simulations shown in Fig.~\ref{ACEXF} shows
good agreement with expectations. The lower amplitude of the reflected pulse is likely
due to imperfections in the coupling of the end of the Alumina and the shorting plate.

\begin{figure}[htb!]
 \includegraphics[width=3.5in]{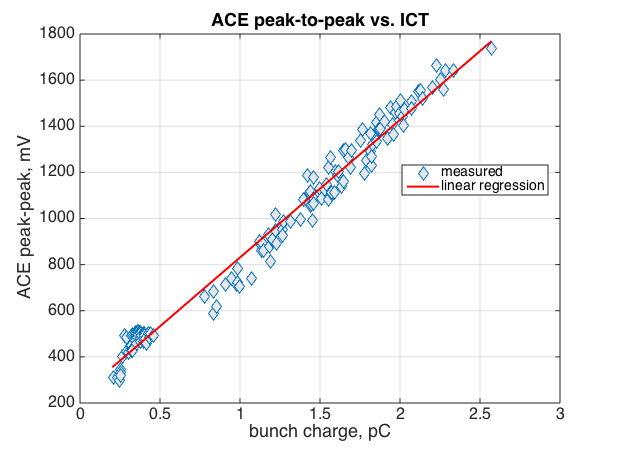}
 \caption{ Scatter plot of ACE field-strength measurements vs. beam currents established from a
 commercial integrating current transformer, showing that the signal response is
 linearly proportional to beam bunch charge.
 \label{ACEvICT}}
 \end{figure}
 
 \begin{figure}[htb!]
 \includegraphics[width=3.5in]{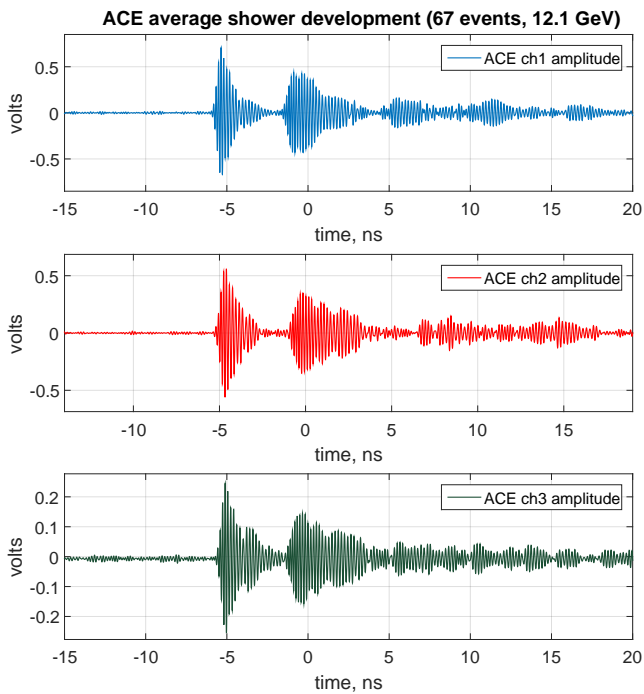} 
 \caption{ Average microwave signals for the 12.16~GeV showers in ACE, for
 67 events of higher SNR. The detector for this run is operating with a full set
 of tungsten pre-shower blocks, at liquid nitrogen temperature, with all low-noise
 amplifiers installed. The figure scales for each subplot are set in the
 relative ratios of (6.1:4.9:2.0) based on the GEANT excess-charge ratios to
 illustrate that the amplitude closely scales with these quantities.
 \label{showerdev}}
\end{figure}

\begin{figure*}[htb!]
 \centerline{\fbox{\includegraphics[width=3.65in]{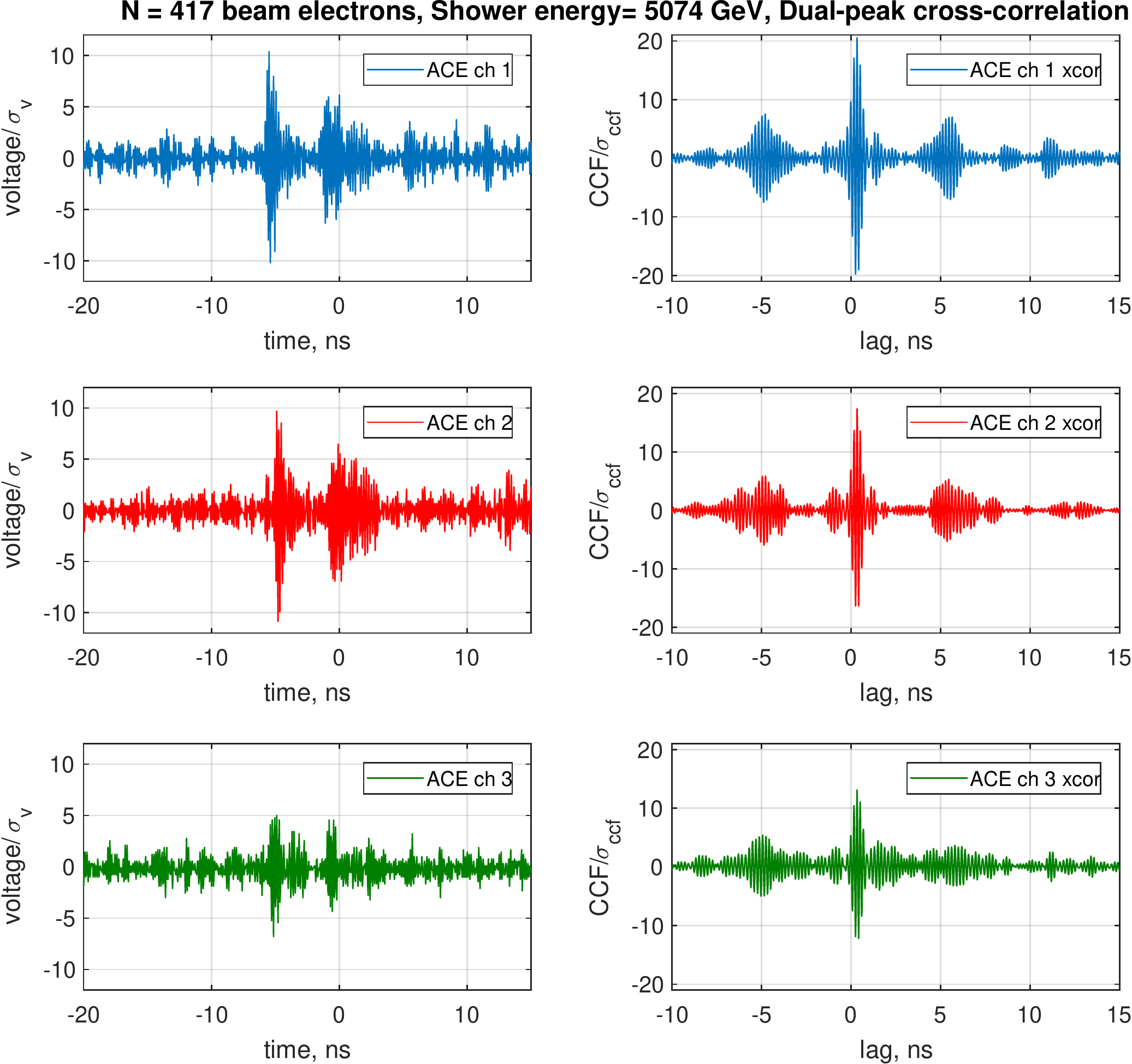}}~\fbox{ \includegraphics[width=3.65in]{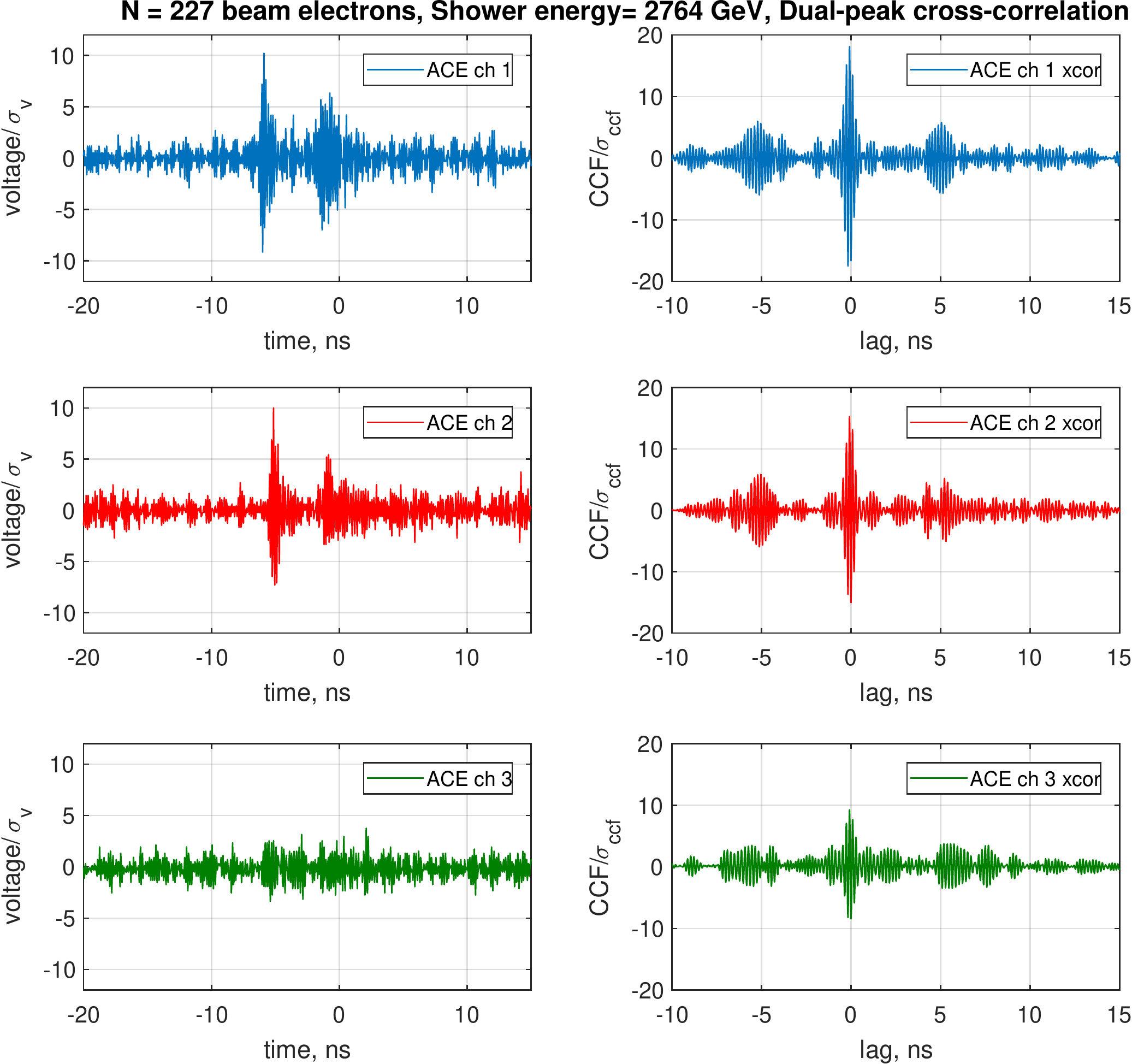}}}
  \centerline{\fbox{\includegraphics[width=3.65in]{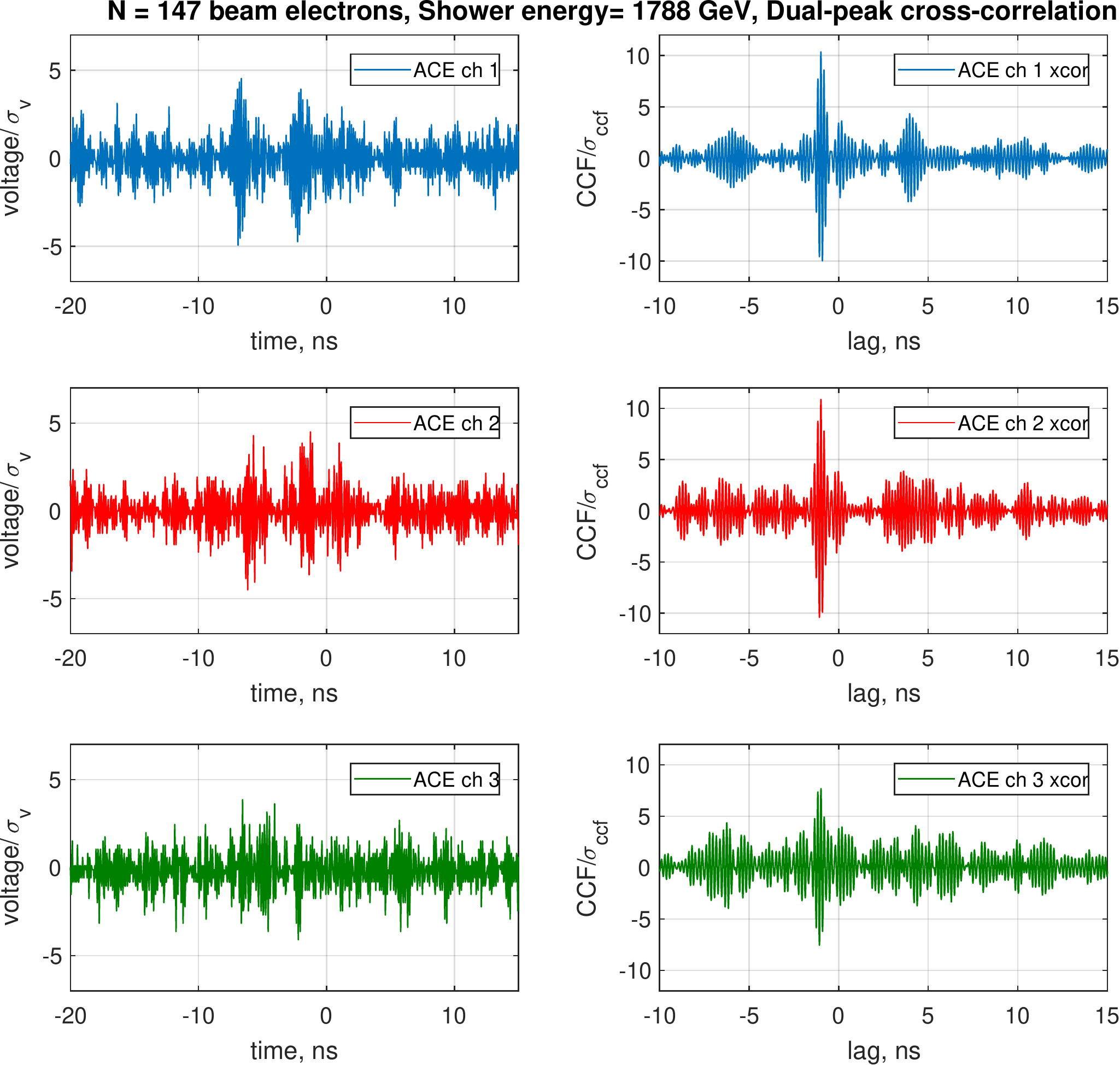}}~\fbox{ \includegraphics[width=3.65in]{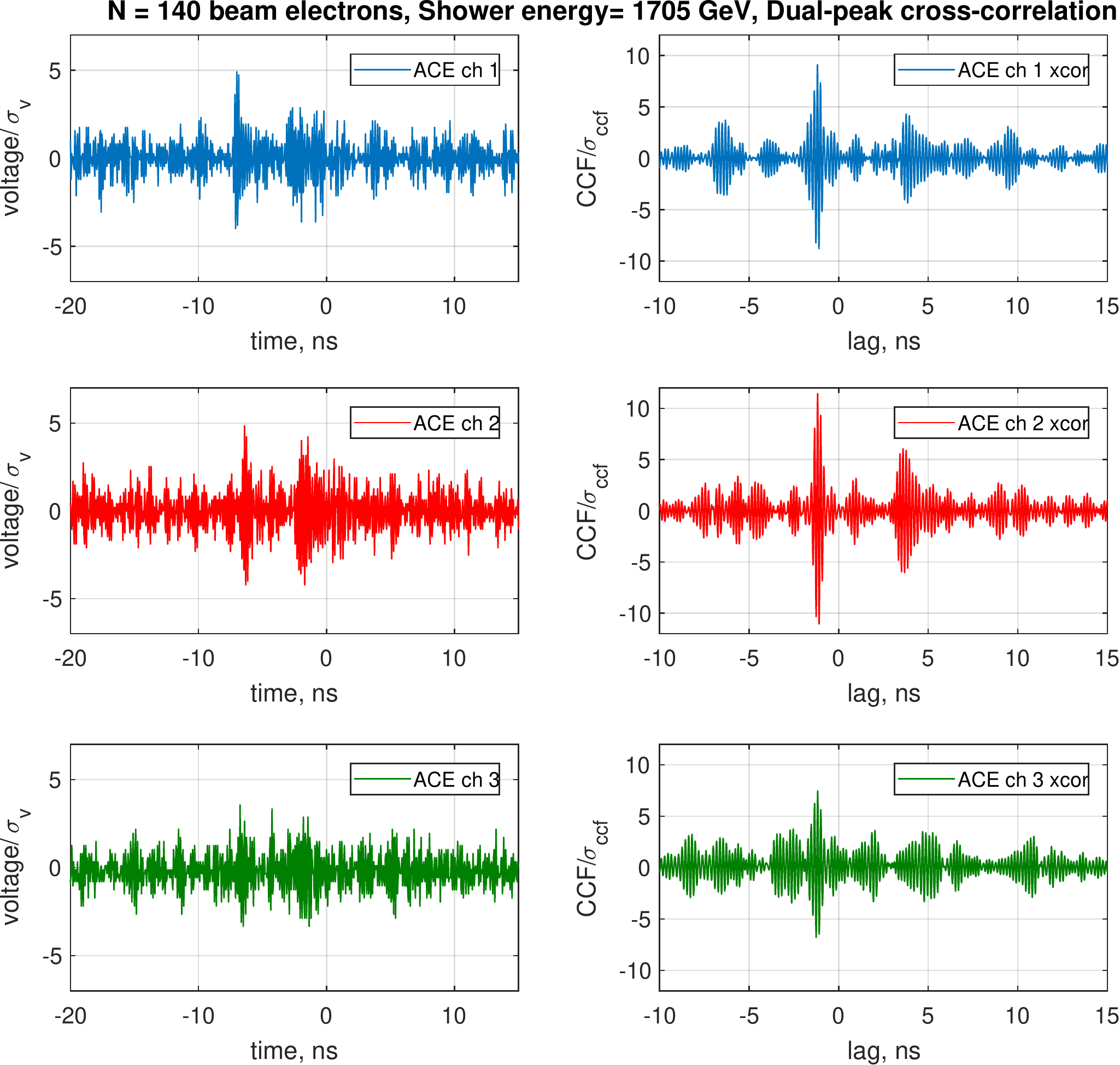}}}
 \caption{ Four typical ACE events at low beam currents, with decreasing total shower energy from upper left
 to lower right. On the left in each set of plots is the measured voltage from each element, normalized to 
 the thermal noise RMS; on the right is the template cross-correlation function, also normalized to its
 RMS noise level. Approximate time offsets are removed from the time axes of the data, giving CCF lags that
 are close to zero, but not required to coincide with zero.
 \label{Acev1ev1}}
 \end{figure*}
 
 \begin{figure}[htb!]
 \includegraphics[width=3.75in]{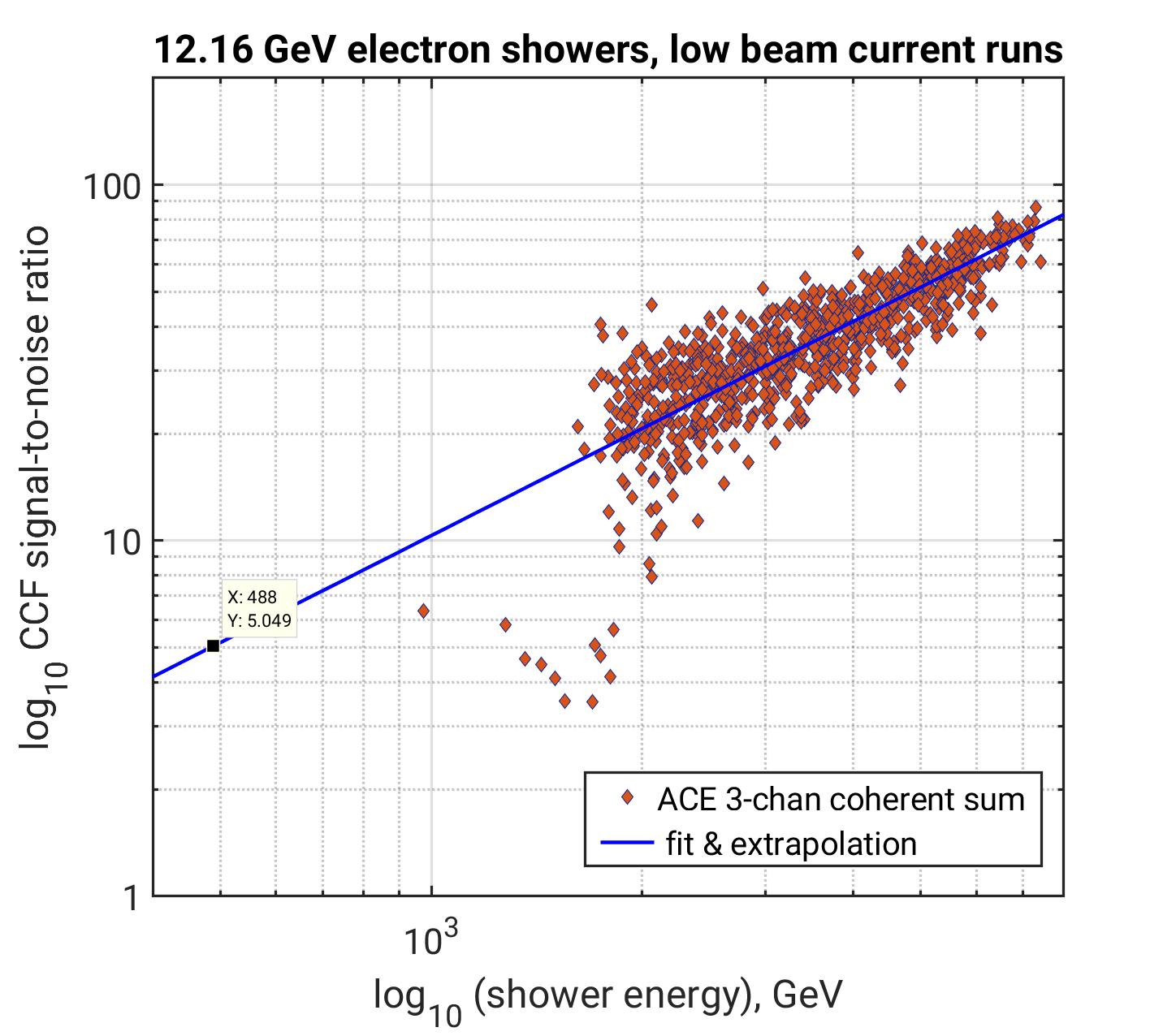} 
 \caption{Scatter plot of the SNR of the phase-aligned sum of the CCFs for all three ACE elements,
 along with a fit and extrapolation to the threshold energy.
 \label{Xcscat}}
\end{figure}

Fig.~\ref{ACEvICT} shows the correlation of the peak-to-peak amplitude of the ACE signals with
beam current measurements from a commercial Integrating Current Transformer (ICT) used
to accurately estimate the actual delivered beam current to our target. This covers a range
from where the ICT drops below its limiting sensitivity at about 0.2 pC per bunch, up to
about an order of magnitude higher. Slight deviations from a linear trend could be due to 
uncalibrated ICT or ACE non-linearities, but overall the correlation is quite good, indicating that the emission
in ACE is dominated by coherent effects, as expected.

\subsection{Low beam current runs}

Since one of our goals in T-530 was to operate the detector at or near its threshold of least-count
energy sensitivity, we used a combination of beam collimators, screens, and a momentum slit to
reduce the beam current to the minimum level where the signal could still be detected, now
using our full low-noise amplification signal chain, and liquid nitrogen to reduce the thermal noise
of the system. In addition, because we wished to enhance the Cherenkov 
emission from the Askaryan effect well above any 
subdominant transition radiation effects, we focused our efforts on using higher energy bunches,
in this case 12.16~GeV which was available during our run. With the tungsten blocks
in place, the ACE counters measure the shower at depths of between 4-8 radiation lengths.

To trigger the system and independently establish the beam current we used a separate, thin-target optical Cherenkov
detector coupled to a silicon-PMT (SiPMT) array. In practice, we found that the SiPMT detector
was usable down to a minimum beam current of $\sim 120-140$ electrons per bunch, and thus for
this experiment the SiPMT determined the the detection floor, equivalent to a 
least count energy of order 1500~GeV per bunch for the 12~GeV runs. 
The composite bunch energy is our effective proxy
for the single particle energy of a secondary in a vertex collision experiment
(assuming that the bulk of its energy is deposited in an EM shower). 
While our minimum trigger energy was quite high in this experiment, it afforded us with 
sufficient data to estimate the intrinsic scaling to lower energies, including
what could be achievable with more aggressive cooling, and higher-order
detector combinatorics. 

To ensure that the behavior of the instrinsic signal scaled closely from
the high beam current runs, we selected events from the 12.16 GeV low-current 
runs with higher SNR, and created an average profile, as shown in Fig.~\ref{showerdev}.
The signal shapes matches those taken at high beam current quite closely.
The amplitude ratios in this case also follow closely the (6.1:4.9:2.0) ratios of excess charge
for (ACE1:ACE2:ACE3) determined from the GEANT4 simulations above. Here we have scaled each y-axis range
according to the GEANT4 excess charge ratios to illustrate this result by the graphical similarity. 
Only ACE3 deviates modestly from the excess-charge ratio, with an amplitude
slightly larger than given by simulation; this may be due to the effect of
the underlying Rician amplitude statistics of a signal in the presence of
thermal noise, since ACE3 has the lowest SNR of the three channels.

Fig.~\ref{Acev1ev1} shows four typical events approaching SiPMT counter minimum energy.
In each of the four events, on the left is the direct scope record, sampled at 20~Gsamples/sec,  
with the three ACE elements in the order
front-to-back appearing top-to-bottom in the figure panes. As noted above the pre-shower tungsten is
about 4 RL thick, each ACE element adds an additional 0.23 RL from the copper waveguide and
alumina, and the later tungsten layers are each 2.5 RL. Thus the shower reaches its maximum
development at these energies between the first and second ACE element.
A reduction in microwave pulse amplitude in the later counters is evident.

On the right hand side of each of the figures, we show the cross-correlation function (CCF) of the signal with
the signal template measured from an average of high-SNR events as shown in Fig.~\ref{showerdev}. 
The CCF is equivalent to using a matched filter based on the impulse and reflection, and yields a correlated amplitude,
which is normalized to the root-mean-square of the CCF at non-signal lags. In effect, the CCF also
removes the dispersion effects of the waveguide, and because the thermal noise is uncorrelated to
the signal pulse, the result shows significant improvement in SNR, exceeding a factor of two in many cases.
The CCF also estimates a time lag for the signal relative to the template, thus providing a high precision measure of the
relative delay between the pulse and template. Both of these parameters are 
representative of the processed data that can be expected from such a detector in practice.

%

Because of the excellent timing precision afforded by the microwave signals, we can 
coherently sum all three of the ACE channels, using phase alignment provided by the CCF,
to get an estimate of the least-count energy achievable with the current system. We define the least-count
energy by requiring at least a $5 \sigma$ detection above thermal noise. 
Fig.~\ref{Xcscat} shows a scatter plot of the combined coherently-summed SNR of
the CCF,  with a fit to allow for extrapolation to the least-count energy,
giving 480~GeV at the $5\sigma$ level in this case. The extrapolation is necessary because the
unexpected high energy threshold of our SiPMT detector. 
The least count energy is dominated in our case by the system thermal noise,
which we estimate to be of order $T_{sys} \simeq 18-25$~K for our LNAs used at liquid nitrogen temperatures.

The estimate from Tamm theory above indicated a single-element $5\sigma$ voltage threshold for ACE1
of about 1000~GeV. Empirically, the template CCF appears to improve the SNR by about a factor of
two, and energy threshold should improve approximately as $1/\sqrt{SNR}$, which leads
to a Tamm theory single-element estimate of order 700~GeV, close to what is observed.
The combination of three elements, with improvement going as $\sim 1/\sqrt{N_{det}}$,
implies a threshold least-count energy of about 400~GeV,  commensurate with the experimental result.
Thus while of necessity we used some {\it ad hoc} assumptions to apply the theory, there is
experimental indication that these are not far from correct.
 
In general, the least-count energy should scale with $\sqrt{T_{sys} / N_{det}}$, thus operating with 
better LNAs, colder cryogens, and more detectors in combination can reduce this by factors of several, possibly
approaching 100~GeV. 

\subsection{Energy calorimetric scaling}

\begin{figure}[htb!]
 \includegraphics[width=3.35in]{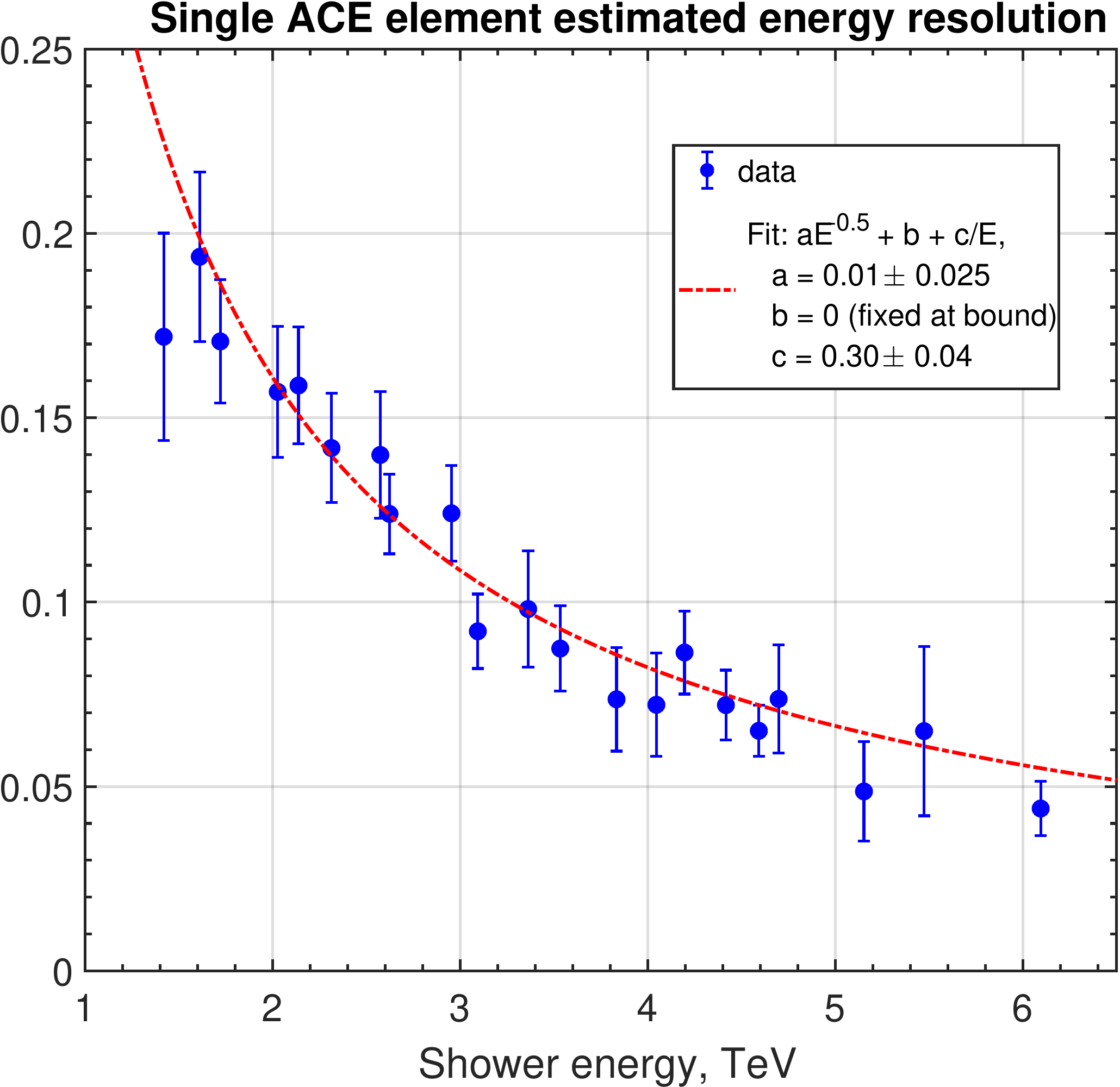}
 \caption{ Energy resolution of a single ACE detector element, estimated against a
 quasi-calibrated second ACE element; thus the errors are a convolution of the
 intrinsic errors of two ACE elements. 
 \label{Eresnew}}
 \end{figure}

While our goal for ACE was not to create a new shower calorimeter, the Askaryan process
by its nature lends itself to calorimetry, and it is thus useful to investigate
the energy response to ensure that we understand the behavior of the system vs. energy.
Fig.~\ref{Eresnew} shows the energy resolution vs. the beam energy. Because the energy
measurement scatter was dominated by the coarse resolution of our SiPMT detector which operated near
threshold, we used a two step process to estimate the energy resolution curve. First the
average energy response function of the SiPMT was used to calibrate a scale for one of the ACE
elements. Then we correlated the quasi-calibrated ACE element against the second ACE element
to get a resolution function that reflects more accurately the quadrature response of the two
ACE elements with respect to one another.

 Fig.~\ref{Eresnew} shows the estimated energy resolution function based on the method
 outlined above. We confirmed that the measured data were dominated by thermal noise, and 
 thus the CCF amplitude errors are uncorrelated between the two. Thus the estimated error bars
 as well as the fractional energy resolution, which arise from the convolution of the errors from
 two uncorrelated detectors, is estimated 
 to be $\sqrt{2}$ larger than what may be achieved intrinsically from a single ACE detector
 element. As noted above, the least count energy is quite high, but to the degree that
 the errors are dominated by thermal noise, the entire graph should scale down with
 a reduction in thermal noise. It is also worth noting that energy resolution 
 is normally quoted for an entire detector system, which in our case would most
 likely improve the resolution by $\sqrt{N}$ for $N$ detector elements.
 
 We also fit these data to a standard parametric curve vs. shower energy $E$, and find that the
 fit is dominated by the $1/E$ term; for these data we find $\Delta E / E = 0.35 E_{TeV}^{-1}$
 for the quadrature response, or $\Delta E / E = 0.23 E_{TeV}^{-1}$ for a single calibrated
 element. 
 
 If in fact we aggressively reduce the thermal noise to put the least count energy down at
 $\sim$100~GeV as described above, the equivalent curve for $N$ samples of the shower would be 
 $$(\Delta E / E)_{scaled} = 0.23 (E/(100~{\rm GeV}))^{-1}N^{-1/2}~.$$ 
 For $N\simeq 4$ which is realistic for
 a typical particle-induced shower, the resulting energy resolution is $0.12/(E/{\rm 100~GeV})$.
 While not of interest for current collider detectors, future $\sim 100$~TeV-scale
 colliders may benefit from this technology for shower calorimetry as well as timing. 
 It is notable also that such technology could be applied immediately to
 heavy ion collisions, where nuclear fragments often carry multi-TeV energies
 in the angular region very close to the beam.
 Current high-rapidity forward collider detectors also do not generally provide high precision timing of such events,
 and thus the timing precision we detail in the following section may provide
 new tools for such investigations.
 
 \subsection{Event timing}
  
 \begin{figure}[htb!]
 \includegraphics[width=3.7in]{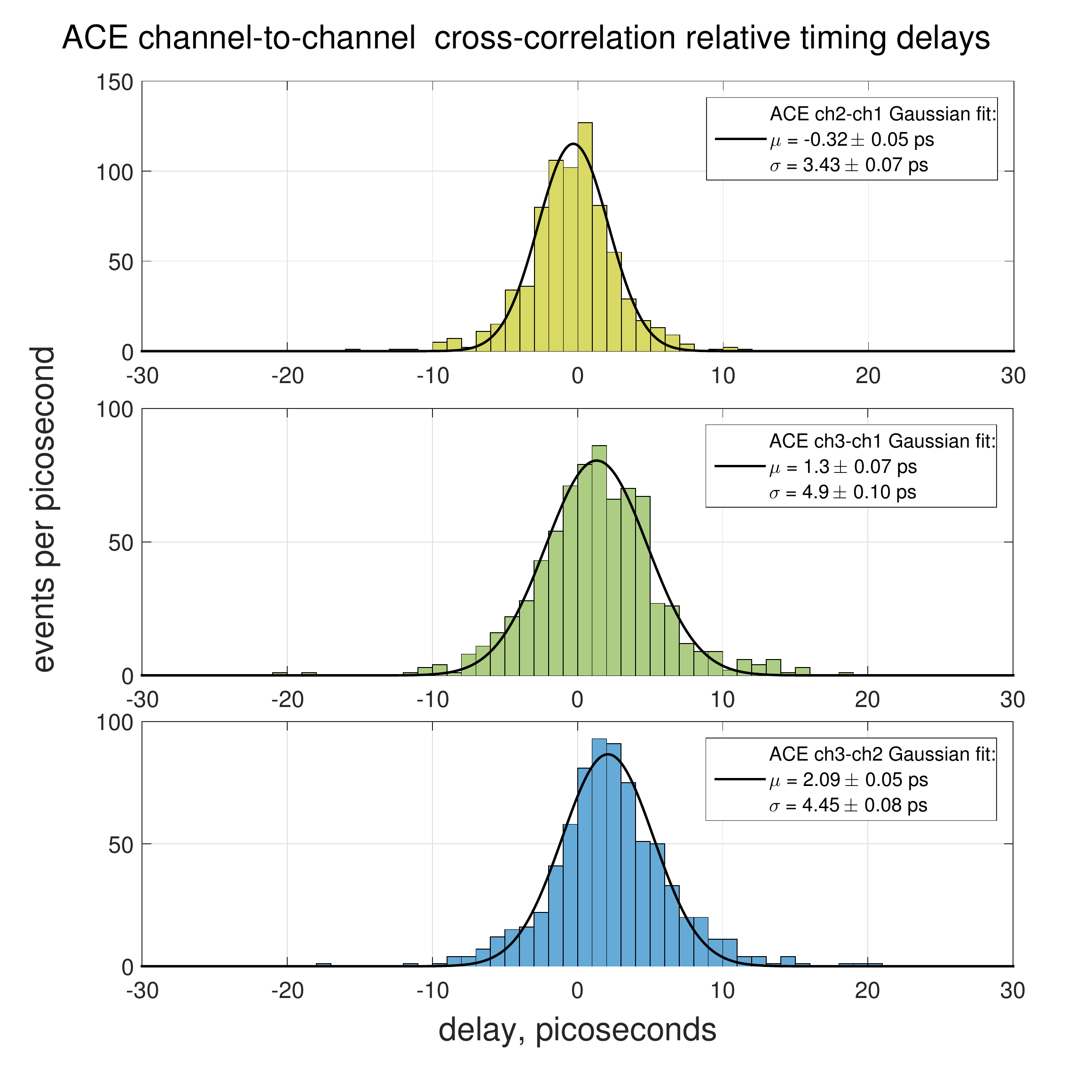}
 \caption{ Relative time resolution of the three ACE detector elements taken in pairs.
 The width of the distribution is thus a convolution of the instrinsic timing resolution
 of the two elements used in each case, increasing the instrinsic width by $\sim \sqrt{2}$. 
 In each case an overall constant approximately
 equal to the channel-to-channel delay has been removed.
 \label{DeltaAllHist}}
 \end{figure}
  \begin{figure}[htb!]
 \includegraphics[width=3.65in]{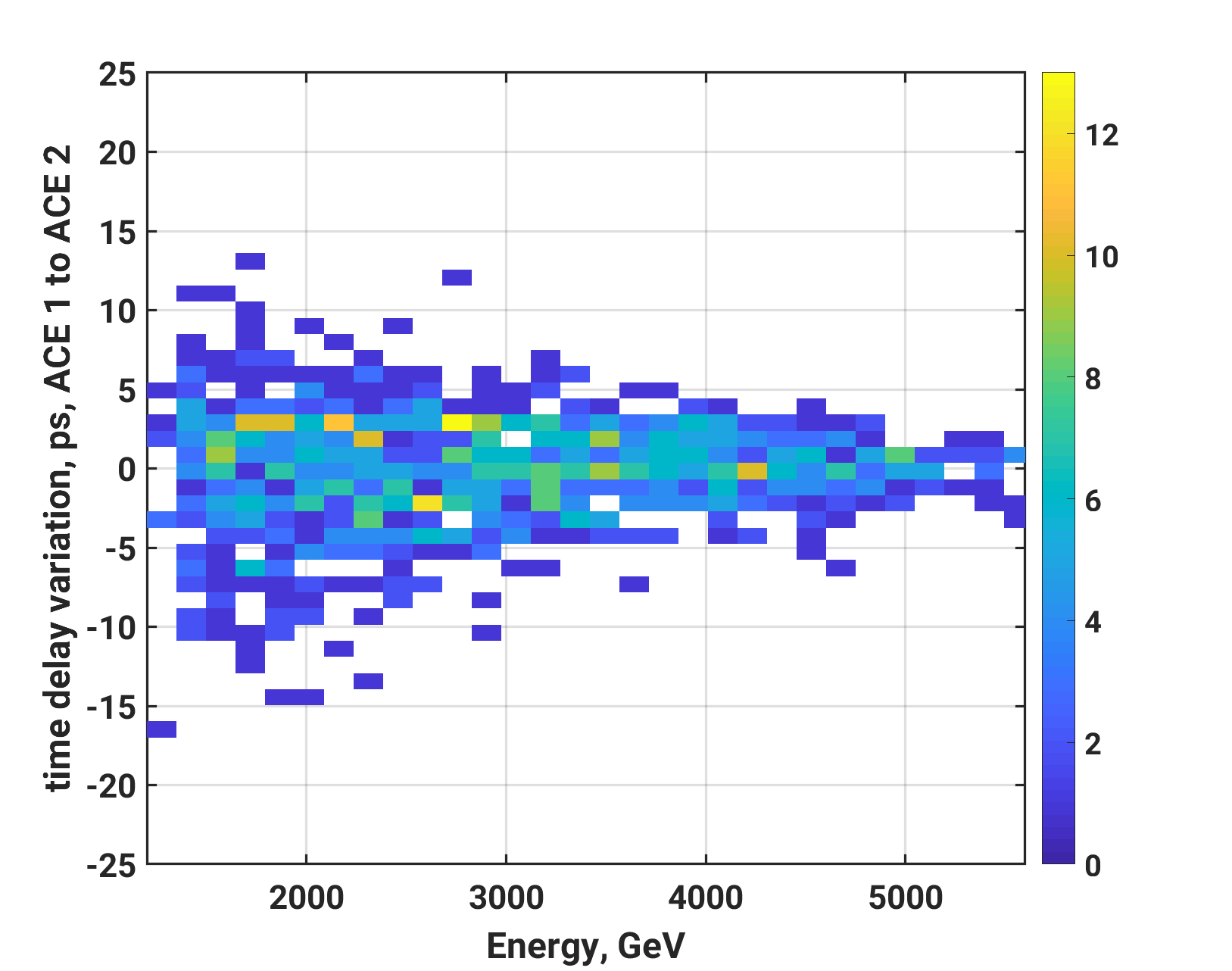}
 \caption{ Time delay resolution as a function of shower energy, for the same data as shown
 in the previous plot. \label{timedelay}}
 \end{figure}

 As we noted above in describing the phasing of the three CCF signals to produce a coherent 
 amplitude estimate, the CCF produced by the template-matched optimal filter used in Fig.~\ref{Acev1ev1}
 above also yields a resolved time of arrival for the pulse. Again, since the time resolution
 precision afforded by the SiPMT detector, with a rise time of order 1~ns, was 
 much coarser that the ACE elements, we estimate relative time differences from one ACE element
 with respect to another measured in all three channels within a single high-bandwidth
 oscilloscope,  which preserves the intrinsically small jitter of the scope in our
 measurements. Since the thermal noise that dominates the errors in our measurements is
 intrinsically uncorrelated between different channels (we have verified this in many such
 experiments), the relative time differences are an accurate measure of the joint 
 statistical precision that is achievable with these devices.

Fig.~\ref{DeltaAllHist} shows the time resolution for the low-beam current data
at 12.16~GeV electron energy, and composite bunch energies in the range of $\sim 1$ up
to about 5~TeV. The fitted precision for measurement between the two higher SNR ACE detectors
is $3.4\pm 0.1$~ps, implying an intrinsic resolution of about 2.4~ps for each element.
Some fraction of the scatter is also due to the intrinsic jitter of the Tektronix
oscilloscope, which specifies a sample-to-sample time difference accuracy of 0.64~ps. 
Removing this from the measured jitter in quadrature gives a slightly better intrinsic
resolution of 2.3~ps for the ACE elements.

Fig.~\ref{timedelay} illustrates the measured time resolution as a function of energy.
These data were taken during a run that was constrained in time and thus the statistics are
not high, but the clear trend toward tighter time resolution with higher shower energy
is evident, following an $E^{-1}$ dependence in a manner similar to the energy resolution.

If we again project these results down to lower energies based on coherent combination of four detectors,
then at liquid nitrogen temperatures, a least-count energy of $\sim 280$~GeV,
and a reference clock with sub-picosecond precision, it is straightforward
to scale our current results to a timing precision of 
\begin{equation}
\sigma_{\Delta t, scaled} = 1.2~{\rm ps}~ \frac{280}{E_{GeV}}~.
\end{equation}
At this level of precision, it is evident that the requirement for precise timing transfer of clock signals to
the digital data acquisition system becomes acute. In addition, high-precision samplers 
with accuracy and resolution comparable to high-end realtime oscilloscopes are also a necessity.

\section{Discussion}

The value of this technology will for many applications depend strongly
on the limiting least-count energy threshold at which a shower, produced as part of
a particle jet from a collision, is detected. Thus the scaling laws that determine this are important to
delineate. We discuss several of the important factors here.

\subsection{System noise temperature}

It is convenient to work with the field strength of the signal and the resulting
induced detector voltage, since in this form the coherent signal strength scales linearly with
the shower energy. The root-mean-square (RMS) detected voltage from a thermal noise source is then
 given by $$V_n = \sqrt{k T_{sys} Z \Delta f}$$ where $k$ is Boltzmann's constant, $Z$ the impedance
 of the receiver, and $\Delta f$ is the frequency bandwidth. The receiver impedance is
 typically $Z=50~\Omega$. The system noise $T_{sys}$ is dominated by the low-noise amplifier,
 but may also contain some contribution due to the loss-tangent of the dielectric, or resistive losses
 in the waveguide walls. For alumina and a high-conductivity-coated and polished waveguide
 surface, the latter contributions are usually negligible.
 
 Since the signal for a coherent source is linearly proportional to beam current and thus shower energy
 for the Askaryan effect, the net effect is that energy threshold improves as
 $\sqrt{(T_{sys}\Delta f)^{-1}}$. Although $Z$ also plays a role here, it is usually constrained by impedance matching
 requirements in the system and is thus not a free parameter. Bandwidth is also constrained in a
 waveguide, with the upper limit governed by the lowest-order mode requirement, and the low end
 limited by the waveguide cutoff. Thus for these practical reasons, lowering the system temperature
 is a straightforward way to improve the energy threshold.
 
%
 In our case, cooling to liquid helium temperatures ($\sim~4.2$~K),  
 LNAs with noise temperatures of order 1-2~K are commercially available.
 If the waveguide and dielectric were also maintained close to liquid helium temperatures,
 $T_{sys} \leq 2$~K is possible, and a four-element ACE cluster could achieve a least-count shower energy in our
 system of order 100~GeV. In addition, the linear dynamic range of
 these detector elements is limited only by the dynamic range of the amplifier chain.
 
 In typical applications, once the signal has been amplified by $\geq 30$~dB in a
 first, low-noise stage, there is no penalty for splitting the signal to go to
 several different gain stages. The base level thermal noise for $T_{sys} = 20$~K 
 in a 3~GHz band is -91~dBm. For a +30-40~dB LNA which compresses at 0~dBm, the dynamic range
 is already 50-60~dB, a factor of over 300-1000 in shower energy (since the Askaryan field strength
 scales directly with it). At 100 GeV least-count shower energy, the linear dynamic range will approach
 100~TeV in shower energy.
  

\subsection{Track length in detector}

In Tamm's theory, the track length $L$ plays a complex role in the resulting intensity of the emission,
appearing both as a scale factor on the overall power, and as a factor in the resulting angular distribution
of the radiation via the $sinc$-like term. In our case the requirement for coupling to the waveguide
dominant mode places constraints on the range of track lengths that are possible. For a beam
entering perpendicular to the waveguide, the track length is fixed at the short dimension of the
waveguide, 6.3~mm in our case. If the waveguide boundary were absent, the emission would
form a cone peaking with polar angle $\sim 70^{\circ}$ from the track. Only a fraction of the solid
angle is directed along the waveguide longitudinal direction. This is reflected in a coupling factor
$\kappa < 1$ for radiation into the waveguide. 

One way that the track length could be increased would be to direct the particle tracks into a more
aligned configuration with the longitudinal axis of the waveguide. An analogous method is used in forward
high-rapidity  calorimeters
which use optical fibers nearly aligned with the beam axis, in the so-called {\it spaghetti calorimeter}
configuration. For optical Cherenkov emission, this geometry trades a long track length in the fiber for
a relatively low efficiency for light coupling to the fiber. A similar approach may in fact yield
lower least count energies for an ACE as well, but we were unable to test these kind of
geometries in the current experiment.

Another way to increase track length in the waveguide element is to move to lower frequencies and
thus larger waveguide cross sections. For example, Alumina-loaded WR-112 would more than double
the track length compared to WR-51; however, the usable bandwidth would be halved, and the net
improvement in least-count energy would be at best $1/\sqrt{2}$, but with potentially a loss in
timing resolution. In fact if the shower 
microwave coherence obtains up to 8 GHz, there might be no improvement in the threshold for
WR-112 compared to WR-51, because the microwave Cherenkov signal grows with frequency,
only flattening out once the shower Moliere radius approaches the wavelength of the
microwave signal. Our studies here did not explore this parameter space; optimization will depend
on exactly which measurements are of most interest for a specific investigation.

A more direct way to increase length by a factor of two for a given rectangular waveguide is
to use waveguide with a {\it square}, rather than standard rectangular, cross section. 
Square waveguide support two degenerate but orthogonal lowest order modes,
the $TE_{10}$ and $TE_{01}$ modes. The degeneracy means that both modes can and will be present
in the same frequency range, but because of orthogonality, they are not necessarily excited
together, especially if the current element that excites the waveguide is aligned with
one of the modes and not the other, as is the case for a shower propagating perpendicular to
one of the waveguide faces. Square waveguide also supports lower intrinsic ohmic losses.
In effect, the two degenerate modes can be thought of as two crossed linear polarizations
in the waveguide. It is possible to then couple them out independently as such.
We have explored this option only in preliminary fashion for this study, but it appears
a promising approach, which leads directly to a factor of two reduction in the least-count 
energy, placing the least count energy well below 100~GeV for aggressively cooled detectors.

\subsection{Magnetic field effects}

In our measurements, no magnetic field was applied to the system. In a collider detector, 
magnetic fields up to several Tesla or more may be present in the system, to provide
charge separation in tracking instruments. It is thus important to consider the effects of such
ambient fields on an ACE. 

While the presence of a strong field will have no effect on the alumina or copper, both of
which have negligible permeability, the field could have negative effects on a low-noise
amplifier. Such effects have been measured~\cite{Daw97}: in the case where a 3.6T field
was oriented perpendicular to the plane of an LNA, an increase of noise figure was observed
from 5.5K to 9K at liquid helium temperatures. Fortunately, such effects can be mitigated by
orienting the plane of the amplifier with the field, and in this case the effects were observed
to vanish~\cite{Daw97}. Such an orientation would in any case be natural for the
field configuration used in typical collider detectors.

Magnetic effects on the shower development are more complex to consider. In our experiment
with 4RL of tungsten pre-shower, and assuming a 50~GeV primary,
the mean shower energy of the electrons entering our
waveguide would still be above 1~GeV, and even a 10T field would give a gyroradius of
more than 30~cm, leading to very little charge separation on the scale of the waveguide. However,
in that case, the shower would enter the waveguide only partially developed, well 
below its maximum, and thus likely to produce a lower amplitude.
Shower maximum would occur at roughly 12-14 RL of tungsten, and if this depth of
pre-shower was used, the mean electron energy would be of order the critical energy in tungsten,
around 8~MeV, with a corresponding gyroradius of $2.8~(\rm{B}/{\rm T})^{-1}$~cm.
Any significant field of a few Tesla or more will lead to significant charge separation 
as the shower transits the detector. This could lead to increased amplitude, since the full charge,
rather than the charge excess, may come in to play in generating the signal, but
the effects will also complicate the calibration of the system, and will require detailed study.
Optimizing the design of the pre-shower system will thus require careful thought and simulation. 

\subsection{Applications.}

As we have stated above, the most straightforward practical value for an ACE-like instrument is in providing one or more
timing planes in larger hybrid detector.  ACE elements could occupy discrete layers
at several depths in either an electromagnetic or hadronic calorimeter, taking advantage of the existing
absorbing layers to sample showers at various depths. 

For a timing plane, our results indicate that the best resolution
will be obtained if the detectors sample showers near their maximum development. 
For this case there is no reason to separate the detectors along the shower direction
so that they sample different shower depths; for improved resolution they should be
arranged back-to-back to provide uncorrelated measurements at close to the same shower
depth.
A timing layer of four loaded WR51 waveguides would require a thickness of 33 mm for the detectors,
with a column density of about 1 radiation length, 
along with possible several thin-walled stainless-steel layers assuming the timing plane
would require its own dewar for cooling. If the timing layer is embedded in a detector
which is already cryogenically cooled, then the dewar would not be necessary,
and the heat load from the ACE elements is very small, tens of mW per channel or less.

Several layers could then be used to provide better coverage of showers developing deeper
in the overall detector system. Our ACE results apply directly to high-energy photon and electron/positron showers,
but large subshowers of a hadronic interaction would also be detectable above the
energy threshold.

\section{Conclusions}

We have demonstrated that the coherent microwave Cherenkov radiation produced by the Askaryan effect
can provide a methodology for precise measurements
of both the energy and arrival time for electromagnetic showers produced by hadronic or
electromagnetic interactions of high-energy particles, in a bounded, dielectric-loaded waveguide.
We find that the microwave impulses produced by transit of a secondary EM shower
through an Alumina-loaded waveguide can be timed to a precision of a few picoseconds or better,
once the impulse amplitude exceeds the system thermal noise by a factor of 2-3. 

While the several hundred GeV least-count energy of the current experiment is too high to
be of immediate relevance to current collider detectors, improvements in detector cooling
based on commercially available low-noise amplifiers and liquid helium cryogens,
along with more favorable detector geometry, 
could lead to sub-100~GeV thresholds, with exceedingly high dynamic ranges, up to 100~TeV per shower.
The resulting radiation-hard detector elements
could provide planar sections in future large composite detector systems, and would provide 
picosecond timing of a subset of particles in a collision, as well as complementary calorimetric
information as well. Such detector planes are potential candidates for technologies in next-generation
colliders, as well possible augmentations to heavy-ion detectors even in their current form.

We thank the excellent staff at the SLAC National Accelerator Laboratory for their
support of this project. This material is based upon work supported by the Department
of Energy under Award Numbers DE-SC0009937, DE-SC0010504, and DE-AC02-76SF0051.


\begin{thebibliography}{99}

\bibitem{TeVColliders} M. L. Mangano, editor, Physics at the FCC-hh, a 100 TeV pp collider,
CERN-2017-003-M, ISBN (Print) 978-92-9083-453-3, ISBN (PDF) 978-92-9083-454-0, (2017),
DOI: \url{ http://dx.doi.org/10.23731/CYRM-2017-003}.


\bibitem{Chekanov17} S.V. Chekanov, M. Beydler, A.V. Kotwal, L. Gray, S. Sen, N.V. Tran, S.-S. Yu, J. Zuzelski,
Initial performance studies of a general-purpose detector for multi-TeV physics at a 100 TeV pp collider,
JINST 12 (2017) P06009.

\bibitem{White14} S. White,
R\&D for a Dedicated Fast Timing Layer in the CMS Endcap Upgrade,
Acta Physica Polonica B vol. 7, no. 4, 735, (2014).

\bibitem{Albrow14} M.G. Albrow,
Fast Timing Detectors for Leading Protons at the LHC: QUARTIC,
Acta Physica Polonica B vol. 7, no. 4, 719, (2014).

\bibitem{Apresyan17}
A. Apresyan, Precision timing at CMS for HL-LHC, Proc.
of the 12th Trento workshop on advanced silicon radiation detectors (TREDI) 2017 (in press).

\bibitem{Cartiglia17} N. Cartiglia, et al., 
Beam test results of a 16 ps timing system based on ultra-fast silicon detectors,
NIM A850, 1 April 2017, pages 83-88.

\bibitem{Ask62} Askaryan GA,
Excess negative charge of an electron-photon shower and its coherent radio emission,
SOVIET PHYSICS JETP-USSR 14 (2): 441-443 1962;
Askaryan GA,
Coherent radio emission from cosmic showers in air and in dense media,
SOVIET PHYSICS JETP-USSR 21 (3): 658, 1965.

\bibitem{Note1} Because our experiment centers around the investigation of a detector element, the "E" in our acronym will be
used interchangeably for {\it experiment} or  {\it element} where appropriate.

\bibitem{GEANT}  J. Allison , M. Gayer , et. al. ,
Nuclear Instruments and Methods in Physics Research A 835 (2016) 186-225.


\bibitem{XF7} REMCOM XFDTD, XF7 release, \url{www.remcom.com}, 315 South Allen Street, Suite 416
State College, PA 16801.

\bibitem{SLAC01} David Saltzberg, Peter Gorham, Dieter Walz, Clive Field, Richard Iverson, 
Allen Odian, George Resch, Paul Schoessow, and Dawn Williams,
Observation of the Askaryan Effect: Coherent Microwave Cherenkov Emission from Charge Asymmetry in High-Energy Particle Cascades,
Phys. Rev. Lett. 86, 2802 (2001).

\bibitem{Hankins} T. H. Hankins  R. D. Ekers  J. D. O'Sullivan,
A search for lunar radio Cherenkov emission from high-energy neutrinos,
Monthly Notices of the Royal Astronomical Society, Volume 283, Issue 3,(1996), 1027.

\bibitem{GLUE} P. W. Gorham, C. L. Hebert, K. M. Liewer, C. J. Naudet, D. Saltzberg, and D. Williams,
Experimental Limit on the Cosmic Diffuse Ultrahigh Energy Neutrino Flux,
Phys. Rev. Lett. 93, 041101, (2004).

\bibitem{RICE} I. Kravchenko, G.M. Frichter, T. Miller, L. Piccirillo, D.
 Seckel,  G.M. Spiczak, J. Adams, S. Seunarine, C. Allen, A. Bean ,
 D. Besson, D.J. Box, R. Buniy, J. Drees, D. McKay, J. Meyers, 
L. Perry, J. Ralston, S. Razzaque, and D.W. Schmitz,
Limits on the ultra-high energy electron neutrino flux from the RICE experiment,
Astroparticle Physics vol. 20, no. 2, 195, (2003).

\bibitem{ANITA}  P. W. Gorham et al. (ANITA Collaboration), 
New Limits on the Ultrahigh Energy Cosmic Neutrino Flux from the ANITA Experiment,
Phys. Rev. Lett. 103, 051103, (2009).


\bibitem{Tak2000} Takahashi T., Shibata Y., Ishi K., Ikezawa M., Oyamada M., Kondo Y.,
Observation of coherent cerenkov radiation from a solid dielectric with short bunches of electrons,
Phys Rev E. 62(B), 8606 (2000).

\bibitem{cohRad1} Ishi K, Shibata Y, Takahashi T, Hasebe S, Ikezawa M, Takami K, Matsuyama T, Kobayashi K, Fujita Y,
Observation of coherent Smith-Purcell radiation from short-bunched electrons,
Phys Rev E Stat Phys Plasmas Fluids Relat Interdiscip Topics. 1995 Jun;51(6):R5212-R5215;

\bibitem{cohRad2}
Peter W. Gorham, David P. Saltzberg, Paul Schoessow, Wei Gai, John G. Power, Richard Konecny, and M. E. Conde,
Radio-frequency measurements of coherent transition and Cherenkov radiation: Implications for high-energy neutrino detection,
Phys. Rev. E 62, 8590, (2000).

\bibitem{Tamm39} I.E. Tamm, J. Phys. Moscow I, 439, (1939).

\bibitem{Al1} L. W. Hobbs, F. W. Clinard, Jr, S. J. Zinkle, and R. C. Ewing, 
Radiation effects in ceramics, 
Journal of Nuclear Materials 216, (1994), 291-321.

\bibitem{Molla95}
J. Molla, A. Ibarra, and E.R. Hodgson, In-beam dielectric properties of alumina,
Journal of Nuclear Materials, Volume 219, 2 March 1995, Pages 182-189.

\bibitem{Warman91}
John M. Warman et al
Electronic processes in semiconductor materials studied by nanosecond time-resolved microwave conductivity. Al2O3, MgO and TiO2 powders
International Journal of Radiation Applications and Instrumentation. Part C. Radiation Physics and Chemistry
Volume 37, Issue 3, 1991, Pages 433-442

\bibitem{Daw97} E. Daw, and R. F. Bradley, Effect of high magnetic  fields on the noise temperature of a heterostructure
field-effect transistor low-noise amplifier, J. Appl. Phys. vol. 82, no. 4, 1925, (1997). 

\bibitem{Zm03} J. Zmuidzinas,
Thermal noise and correlation in photon detection,
 Appl. Optics vol. 42, no. 25, 4989 (2003).

\end{thebibliography}
\end{document}